\documentclass[aps,pra,onecolumn,amssymb,amsfonts,floatfix,nofootinbib,superscriptaddress,longbibliography,notitlepage]{revtex4-1}
\pdfoutput=1
\usepackage[utf8]{inputenc}
\usepackage{color}
\usepackage{multirow}
\usepackage{dcolumn}
\usepackage{float}
\usepackage{xcolor}
\usepackage{amsmath,amsthm,amsfonts,amssymb}
\usepackage{times}
\usepackage[normalem]{ulem}
\usepackage{physics}
\usepackage{commath}
\usepackage{bm}
\usepackage[pdftex]{graphicx}
\usepackage[colorlinks=false]{hyperref}

\begin{document}

\title{Quantum scarring in a spin-boson system: fundamental families of periodic orbits}

\author{Sa\'ul Pilatowsky-Cameo}
\author{David Villase\~nor}
\affiliation{Instituto de Ciencias Nucleares, Universidad Nacional Aut\'onoma de M\'exico, Apdo. Postal 70-543, C.P. 04510  CDMX, Mexico}
\author{Miguel~A.~Bastarrachea-Magnani}
\affiliation{Department of Physics and Astronomy, Aarhus University, Ny Munkegade, DK-8000 Aarhus C, Denmark}
\author{Sergio Lerma-Hern\'andez}
\affiliation{Facultad de F\'isica, Universidad Veracruzana, Circuito Aguirre Beltr\'an s/n, Xalapa, Veracruz 91000, Mexico}
\author{Lea~F.~Santos} 
\affiliation{Department of Physics, Yeshiva University, New York, New York 10016, USA}
\author{Jorge G. Hirsch}
\affiliation{Instituto de Ciencias Nucleares, Universidad Nacional Aut\'onoma de M\'exico, Apdo. Postal 70-543, C.P. 04510  CDMX, Mexico}

   
\begin{abstract}
As the name indicates, a periodic orbit is a solution for a dynamical system that repeats itself in time. In the regular regime, periodic orbits are stable, while in the chaotic regime, they become unstable. The presence of unstable periodic orbits is directly associated with the phenomenon of quantum scarring, which restricts the degree of delocalization of the eigenstates and leads to revivals in the dynamics. Here, we study the Dicke model in the superradiant phase and identify two sets of fundamental periodic orbits. This experimentally realizable atom-photon model is regular at low energies and chaotic at high energies. We study the effects of the periodic orbits in the structure of the eigenstates in both regular and chaotic regimes and obtain their quantized energies. We also introduce a measure to quantify how much scarred an eigenstate gets by each family of periodic orbits and compare the dynamics of initial coherent states close  and away from those orbits.
\end{abstract}

\maketitle


\section{INTRODUCTION}
\label{sec:Introduction}
The eigenstates of quantum systems that are fully chaotic in the classical limit may not be completely delocalized in the energy shell. This realization~\cite{Heller1984} came as a surprise, since in the classical domain the trajectories do fill the phase space densely. The presence of unstable periodic orbits (UPOs) plays a central role in the phase-space representation of the eigenstates~\cite{Berry1989}. Despite having measure zero, UPOs give rise to quantum scars, which are characterized by enhanced probabilities along the phase-space region occupied by those orbits and thus prevent the uniform distribution of the eigenstates~\cite{Heller1991}. Quantum scars were first studied in one-body systems~\cite{McDonald1979,StockmannBook,Wintgen1989,Ariano1992,Heller1987,Bogomolny1988,Agam1993,Bohigas1993,Muller1994,Kaplan1998,Kaplan1999,
Wisniacki2006,Porter2017} and then in two-dimensional harmonic oscillators~\cite{Keski2019,KeskiHeller2019} and  the Dicke model~\cite{Dicke1954,Deaguiar1992,Bakemeier2013}. More recently, they have been connected also with some special non-thermal states of many-body quantum systems~\cite{Turner2018,Turner2020Arxiv}, although in this case the analysis of the classical limit is still missing.

The present work focuses on the effects of quantum scars in the Dicke model. This spin-boson model consists of ${\cal N}$ atoms collectively coupled with a quantized field. It was introduced to understand the superradiance phenomenon in light-matter systems~\cite{Dicke1954,Hepp1973a,Hepp1973b,Wang1973,Emary2003,Garraway2011} and  it was later used in studies of nonequilibrium dynamics~\cite{Fernandez2011,Altland2012PRL,Shen2017,Lerma2018,Lerma2019,Kloc2018,Kirton2019,Villasenor2020}, including the evolution of out-of-time-ordered correlators~\cite{Chavez2019,Lewis-Swan2019,Pilatowsky2020}, and as a paradigm of the ultra-strong coupling regime in several systems~\cite{DeBernardis2018,Kockum2019,FornDiaz2019,LeBoite2020}. The model has been studied experimentally with cavity assisted Raman transitions~\cite{Baden2014,Zhang2018}, trapped ions~\cite{Cohn2018,Safavi2018}, and  circuit quantum electrodynamics~\cite{Jaako2016}. It has two degrees of freedom and displays both regular and chaotic behavior~\cite{Lewenkopf1991,Emary2003PRL,Emary2003,Bastarrachea2014b,Bastarrachea2015,Bastarrachea2016PRE,Chavez2016} depending on  the Hamiltonian parameters and excitation energy. Scarring in the Dicke model was studied for a low number of atoms in~\cite{Deaguiar1991,Deaguiar1992}, where an algorithm to identify the classical periodic orbits (POs) was implemented. The technique allowed to match the phase-space probability accumulations of the eigenstates with specific POs. The comparison between quantum and classical phase-space distributions was later extended to a large number of atoms in~\cite{Bakemeier2013}. 
             
Recently, in \cite{PilatowskyARXIV} we showed that even in the chaotic high-energy region, all eigenstates of the Dicke model accumulate around portions of the phase-space energy shells of the corresponding eigenenergies, covering at most half of the available phase space. This was done using a phase-space localization measure similar to the one studied in Ref. ~\cite{Wang2020}. 

In this work we identify the two fundamental families of POs that emanate from the two normal modes of the ground-state configuration and study extensively their influence in the phase-space localization of the eigenstates. The orbits change from stable in the regular low-energy regime of the model to unstable as one approaches the chaotic high-energy region. We introduce a measure that quantifies the degree of scarring of an eigenstate by a specific PO and use it to select the states concentrated around those two families of POs. We also show that the energies of these states follow the Bohr-Sommerfeld quantization rules for both stable and unstable POs.

The second part of the paper is dedicated to the dynamics of the Dicke model in the chaotic regime. The evolution  depends on the proximity of the initial state to the POs. In the classical limit, an initial configuration launched exactly at a UPO will remain over it for an infinitely long time. However, the UPOs form a set of measure zero in the phase space, so their existence does not break the ergodic properties of the whole system and initial configurations picked up at random pass arbitrarily close to all other accessible configurations at the same energy~\cite{footExcept}. In the quantum domain, the effects of the UPOs are more dramatic. An initial state defined over a phase space region that includes a short-period UPO should exhibit revivals and, after long times, it should be more likely to be found in the vicinity of that same UPO. Therefore, the state explores an effective reduced volume of the phase space~\cite{Villasenor2020} and displays what is known as a dynamical scar \cite{Tomiya2019} in its infinite-time average. In contrast, initial states that are away from short-period UPOs should follow a path to equilibrium according to random matrix theory, maximally covering the phase space at long times. Since we have two families of POs, we can choose initial coherent states that are very close to them and verify these expectations. 

The paper is organized as follows. In Sec.~\ref{sec:DickeModel}, we present the Dicke Hamiltonian, its properties, and its classical limit. In Sec.~\ref{sec:Scarring}, we identify the families of classical POs emanating from the ground-state configuration and introduce a measure to determine the degree of scarring of the eigenstates. In Sec.~\ref{sec:dynamics}, we analyze the effects of scarring in the dynamics of initial coherent states close to those POs. Our conclusions are given in Sec.~\ref{sec:conclusions}.


\section{DICKE MODEL}
\label{sec:DickeModel}

The Dicke Hamiltonian~\cite{Dicke1954} is given by
\begin{equation}
\label{eqn:qua_hamiltonian}
\hat{H}_{D}=\omega\hat{a}^{\dagger}\hat{a}+\omega_{0}\hat{J}_{z}+\frac{2\gamma}{\sqrt{\mathcal{N}}}\hat{J}_{x}(\hat{a}^{\dagger}+\hat{a}),
\end{equation}
where $\hbar=1$. The model describes a system of $\mathcal{N}$ two-level atoms with atomic transition frequency $\omega_{0}$ interacting with a single mode of the electromagnetic field with radiation frequency $\omega$. The parameter $\gamma$ controls the atom-field coupling, $\hat{a}$ ($\hat{a}^{\dagger}$) is the usual bosonic annihilation (creation) operator of the field mode, and $\hat{J}_{x,y,z}=\frac{1}{2}\sum_{k=1}^{\mathcal{N}}\hat{\sigma}_{x,y,z}^{k}$ are the collective pseudo-spin operators with  $\hat{\sigma}_{x,y,z}$ being the Pauli matrices.

The eigenvalues $j(j+1)$ of the squared total-spin operator $\hat{\bm{J}}^{2}=\hat{J}_{x}^{2}+\hat{J}_{y}^{2}+\hat{J}_{z}^{2}$ determine the different invariant subspaces of the model. We use the symmetric atomic subspace defined by the maximum pseudo-spin value $j=\mathcal{N}/2$, which includes the ground state. The Hamiltonian $\hat{H}_{D}$ commutes with the parity operator $\hat{\Pi}=e^{i\pi\hat{\Lambda}}$, where the operator $\hat{\Lambda}=\hat{a}^{\dagger}\hat{a}+\hat{J}_{z}+j\hat{1}$ has eigenvalues $\lambda=n+m+j$, that correspond to the total number of excitations. The number of photons is given by $n$ and the number of excited atoms by $m+j$, where $m$ is an eigenvalue of the atomic operator $\hat{J}_{z}$.

When $\gamma$ reaches the critical value $\gamma_{c}=\sqrt{\omega\omega_{0}}/2$, the Dicke model presents a second-order quantum phase transition~\cite{Hepp1973a,Hepp1973b,Wang1973,Emary2003}. At this point, it goes from a normal phase ($\gamma<\gamma_c$), where the ground state has no photons and all atoms are in their ground state, to a superradiant phase ($\gamma>\gamma_c$), where the ground state has a finite amount of photons and excited atoms.

\subsection{Classical limit of the Dicke Hamiltonian}

The classical Hamiltonian defined over the four-dimensional phase space $\mathcal{M}$ of the Dicke model, with coordinates $\bm x=(q,p;Q,P)$, is constructed using Glauber-Bloch coherent states~\cite{Deaguiar1991,Deaguiar1992,Bastarrachea2014a,
Bastarrachea2014b,Bastarrachea2015,Chavez2016}
\begin{equation}
\ket{\bm{x}}=\ket{q,p}\otimes\ket{Q,P},
\end{equation}
which are built as a tensor product of Glauber coherent states for the bosonic sector
\begin{equation}
\label{eqn:glauber}
|q,p\rangle=e^{-(j/4)\left(q^{2}+p^{2}\right)}e^{\left[\sqrt{j/2}\left(q+ip\right)\right]\hat{a}^{\dagger}}|0\rangle, 
\end{equation}
and Bloch coherent states for the pseudo-spin sector
\begin{equation}
\label{eqn:bloch}
|Q,P\rangle=\left(1-\frac{Q^{2}+P^{2}}{4}\right)^{j}e^{\left[\left(Q+iP\right)/\sqrt{4-(Q^{2}+P^{2})}\right]\hat{J}_{+}}|j,-j\rangle,
\end{equation}
where $|0\rangle$ is the photon vacuum, and $|j,-j\rangle$ is the state with all atoms in their ground state. The raising (lowering) collective pseudo-spin operator, $\hat{J}_{+}$ ($\hat{J}_{-}$), is defined in the usual way $\hat{J}_{\pm}=\hat{J}_{x}\pm i\hat{J}_{y}$.

Taking the expectation value of the Hamiltonian $\hat{H}_{D}$ under the states $\ket{\bm{x}}$ and dividing it by the pseudo-spin $j$~\cite{Villasenor2020}, we obtain the classical Hamiltonian
\begin{align}
\label{eqn:cla_hamiltonain_QP}
h_\text{cl}(\bm x) & \equiv \frac{\langle\bm{x}|\hat{H}_{D}|\bm{x}\rangle}{j}=\frac{\omega}{2}(q^{2}+p^{2})
 +\frac{\omega_{0}}{2}(Q^{2}+P^{2})+2\gamma Q q\sqrt{1-\frac{Q^{2}+P^{2}}{4}} -\omega_{0} .
\end{align}
We define the rescaled energy corresponding to $h_\text{cl}$ as
\begin{equation}
\label{eqn:norm_ener}
\epsilon=E/j,
\end{equation}
which determines an effective Planck constant $\hbar_{\text{eff}}=1/j$~\cite{Ribeiro2006}.

Depending on the Hamiltonian parameters and excitation energies, different dynamical behaviors of the model are identified, ranging from regularity to chaos. As a case study, we choose $\omega=\omega_0=1$, coupling parameter in the superradiant phase $\gamma=2\gamma_c$, and pseudo-spin value $j=30$ ($\mathcal{N}=60$). For these parameters, the ground-state energy is given by $\epsilon_\text{GS}=-2.125$. The dynamics is regular up to a value of $\epsilon\approx-1.7$ and chaotic for higher energies~\cite{Chavez2016}.


\section{QUANTUM SCARRING}
\label{sec:Scarring}

Quantum scarring is a phenomenon by which some eigenstates of a quantum system get concentrated around the UPOs that appear in the classical limit of the model. To identify the scarred eigenstates of the Dicke model, we must first identify the POs arising in the classical limit.

\subsection{Families of periodic orbits in the classical limit}
\label{sec:UPOS}

\begin{figure}
\centering
\includegraphics*[width=4in]{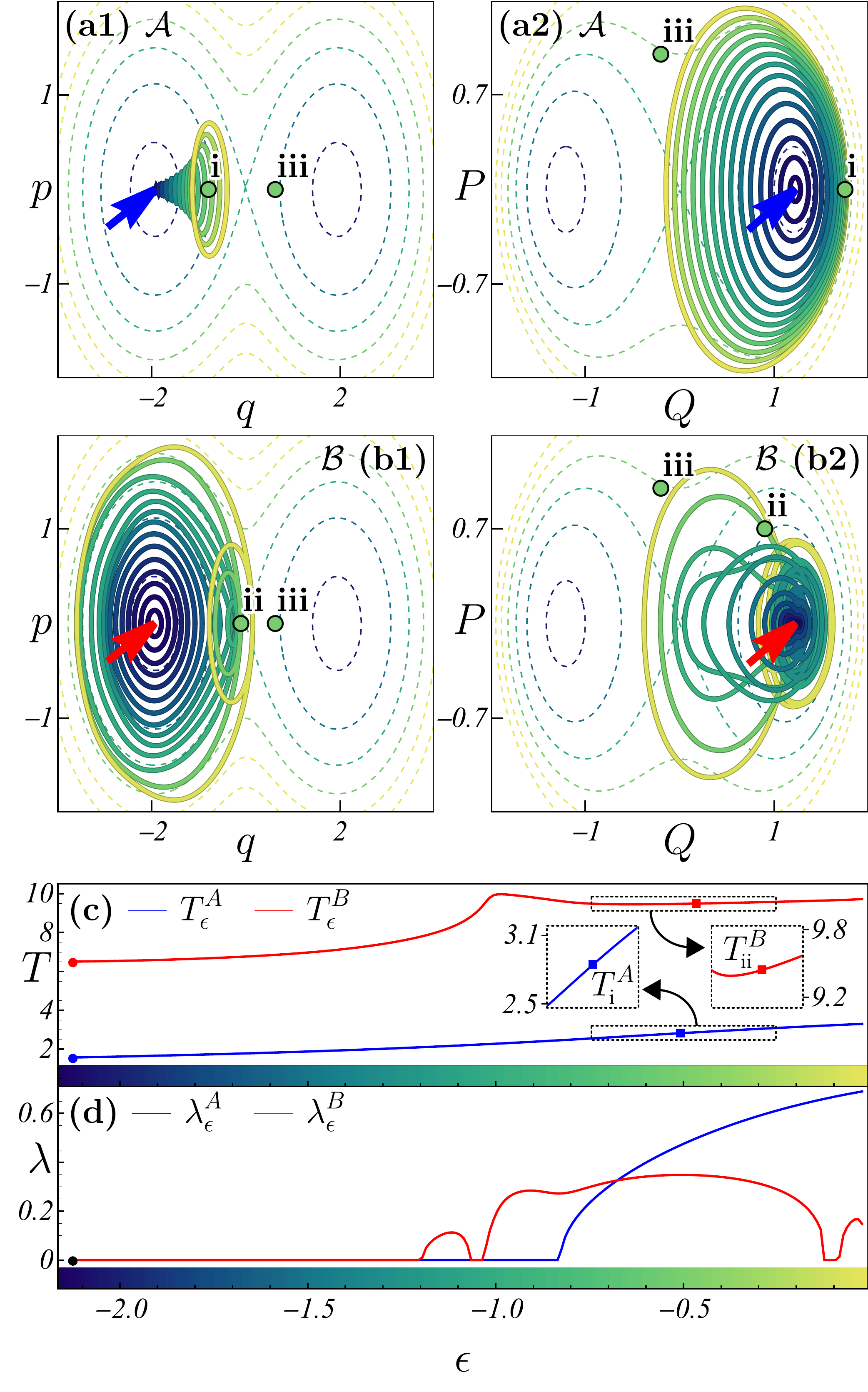}
\vspace{-2em}
\caption{In panels (a1)-(b2): Solid curves are the POs from families $\mathcal{A}$ in (a1)-(a2) and $\mathcal{B}$ in (b1)-(b2) projected into the planes $q$-$p$ in (a1)-(b1) and $Q$-$P$ in (a2)-(b2). Each PO has a different energy. Light colors indicate higher energies, according to the color codes in the horizontal axes of panels (c) and (d). The dashed lines in (a1)-(b2) enclose the available phase-space at different energies. 
The red and blue arrows mark the stationary point $\bm x _\text{GS}$, from which each family of POs emanates. The points marked i, ii and iii are the ones listed in Eq.~\eqref{eqn:CohSta}. 
In panels (c)-(d): The blue and red solid lines represent the period (c) and the maximal Lyapunov exponent (d) of the POs $\mathcal{O}^{A}_\epsilon$ in $\mathcal{A}$ (blue line) and $\mathcal{O}^{B}_\epsilon$ in $\mathcal{B}$ (red line) as a function of energy $\epsilon$. In (c), the blue and red dots at the lowest energy mark the normal periods around the stable stationary point ${\bm x}_\text{GS}$, $T_{\epsilon_\text{GS}}^A$ and $T_{\epsilon_\text{GS}}^B$, respectively. The insets in (c) are zoomed-in plots of $T^A_\epsilon$ (blue) and  $T^B_\epsilon$ (red). The value of $T^A_\text{i}$ [$T^B_\text{ii}$] is indicated with a blue [red] solid square. In (d), the black dot at lowest energy marks the stable stationary point ${\bm x}_\text{GS}$ with zero Lyapunov exponent. 
\vspace{-2em}}
\label{fig:01}
\end{figure}

A PO $\mathcal{O}$  with period $T$ is a subset of the phase space, such that  
\begin{equation}
\mathcal{O}=\left \{\bm x(t) \, \middle\vert \, t\in [0,T] {\hbox {\ \ and\ \ }} \bm x(0)=\bm x(T)  \right \}.
\end{equation} 
The most trivial periodic orbit consists of a single stationary point, where $\bm x_\text{st}(t)=\bm x_\text{st}(0)$ for all times $t$. The classical dynamics given by $h_\text{cl}$ yield several stationary points that may be located by finding  the extrema of the Hamiltonian. For our chosen parameters, there are two stationary points located at the ground state energy $\epsilon_\text{GS}$~\cite{Bastarrachea2014a}:
\begin{equation*}
 \bm x_\text{GS}=\bigg(q=-\sqrt{\frac{4\gamma^2}{\omega^2} - \frac{\omega_0^2}{4\gamma^2}},Q=\sqrt{2-\frac{\omega \omega_0}{2 \gamma^2}},p=P=0 \bigg),
\end{equation*}
and $\widetilde{\bm x}_\text{GS}$, where the signs of $q$ and $Q$ are opposite to those of $\bm x_\text{GS}$. In our case,
\begin{align*}
{\bm x}_\text{GS}=&\big(q=-1.936,Q=1.225,p=P=0\big)&&\text{and}&&
\widetilde{\bm x}_\text{GS}=\big(q=1.936,Q=-1.225,p=P=0\big).
\end{align*}
We first consider $\bm x_\text{GS}$, which is marked with a blue arrow in Figs.~\ref{fig:01}~(a1) and (a2), and with a red arrow in Figs.~\ref{fig:01}~(b1) and (b2).

By considering small displacements around the stationary point $\bm x_\text{GS}$, we obtain two normal frequencies of the system, $\Omega_{\epsilon_{\text{GS}}}^A$ and $\Omega_{\epsilon_{\text{GS}}}^B$, which are given by
\begin{equation*}
\Omega_{\epsilon_{\text{GS}}}^{A,B}=\sqrt{\frac{1}{2\omega^2}\left(\left(16 \gamma^4+\omega^4\right) \pm\sqrt{\left(\omega^4-16 \gamma^4\right)^2+4 \omega^6 \omega_0^2}\right)},
\end{equation*}
where $A$ corresponds to the plus sign and $B$ to the minus sign. For our selection of parameters, $\Omega_{\epsilon_{\text{GS}}}^{A}=4.008$ and $\Omega_{\epsilon_{\text{GS}}}^{B}=0.966$. Correspondingly, we have two normal periods $T_{\epsilon_{\text{GS}}}^{A}=2\pi/ \Omega_{\epsilon_{\text{GS}}}^{A}=1.568$ and $T_{\epsilon_{\text{GS}}}^{B}=2\pi/ \Omega_{\epsilon_{\text{GS}}}^{B}=6.503$.

Let us focus first  on the normal mode of period $T_{\epsilon_{\text{GS}}}^{A}$ with the trivial PO $\mathcal{O}_{\epsilon_\text{GS}}^A=\{\bm x_\text{GS}\}$. By perturbing this stable stationary point and using a monodromy method~\cite{DeAguiar1988} to guarantee the convergence, we can find a new PO $\mathcal{O}_{\epsilon'}^A$ with energy $\epsilon'=\epsilon_\text{GS}+\delta\epsilon$ and period  $T_{\epsilon'}^A=T_{\epsilon_\text{GS}}^A +\delta T$~\cite{Weinstein1973}. This procedure can be successively repeated to increasing energies $\epsilon$, so that a continuous family of periodic orbits $\mathcal{O}^A_{\epsilon}$  is obtained  all the way to the chaotic regime, where the orbits become unstable (see App.~\ref{app:findingPOs} for details). This family of POs is denoted by $\mathcal{A}$,
\begin{equation}
\mathcal{A}=\left \{\mathcal{O}^A_\epsilon \, \middle\vert \, \epsilon_\text{GS} \leq \epsilon \leq 0  \right \} ,
\end{equation}
and it is shown in Figs.~\ref{fig:01}~(a1) and (a2), where the color of each PO indicates its energy (lighter colors mean larger energies). 

We repeat the procedure above for the other normal mode around the ground state with $T_{\epsilon_{\text{GS}}}^{B}$, yielding the POs of family $\mathcal{B}$,
\begin{equation}
\mathcal{B}=\left \{\mathcal{O}^B_\epsilon \, \middle\vert \, \epsilon_\text{GS} \leq \epsilon \leq 0  \right \},
\end{equation}
which is plotted in Figs.~\ref{fig:01}~(b1) and (b2). 

The energy of each PO in the two families identified increases as the POs grow away from $\bm x_\text{GS}$. In Fig.~\ref{fig:01}~(c), we show the period $T^A_\epsilon$ of the POs in $\mathcal{A}$ (blue line) and $T^B_\epsilon$ for the POs in $\mathcal{B}$ (red line) as a function of the energy $\epsilon$. For a given energy $\epsilon$, the period $T^B_\epsilon$ is between three and four times larger than $T^A_\epsilon$, a feature to which we come back to when explaining our next results. In Fig.~\ref{fig:01}~(d), we display the Lyapunov exponents of those POs as a function of energy (see App.~\ref{app:Lyapunovs} for details). The orbits become unstable when the Lyapunov exponents are different from zero. Notice that for family $\mathcal{B}$, the Lyapunov exponent does not grow monotonically with the energy and there are ranges of high energies where the PO can be stable.

The classical Hamiltonian~\eqref{eqn:cla_hamiltonain_QP} is invariant under the transformation $(q,p;Q,P)\mapsto(-q,p;-Q,P)$. This is the classical manifestation of the parity conservation present in the quantum system. The invariance means that if $\mathcal{O}$ is a PO, we may find another PO by mirroring 
\begin{equation}
\begin{gathered}
\mathcal{O}\xrightarrow{\hspace*{1.5cm}}\widetilde{\mathcal{O}}\quad\\
(q,p;Q,P) \mapsto (-q,p;-Q,P), \nonumber
\end{gathered}
\end{equation}
where $\widetilde{\mathcal{O}}$ has the same period, energy, and Lyapunov exponent as $\mathcal{O}$. This transformation yields the mirrored families
\begin{align}
\widetilde{\mathcal{A}}=\left \{\widetilde{\mathcal{O}}^{A}_\epsilon \, \middle \vert \, \mathcal{O}^{A}_\epsilon\in \mathcal{A} \right \} &&\text{and}&& \widetilde{\mathcal{B}}=\left \{\widetilde{\mathcal{O}}^{B}_\epsilon \, \middle \vert \, \mathcal{O}^{B}_\epsilon\in \mathcal{B} \right \}.
\end{align}
The two families $\widetilde{\mathcal{A}}$ and $\widetilde{\mathcal{B}}$ are the ones that emanate from the stationary point $\widetilde{\bm x}_\text{GS}$. 


\subsection{Scarring of energy eigenstates}

The four families $\mathcal{A}$, $\mathcal{B}$, $\widetilde{\mathcal{A}}$, and $\widetilde{\mathcal{B}}$ scar many of the quantum eigenstates, as we show in this subsection.

\subsubsection{Measure of scarring degree}

The Husimi function of a state $\hat{\rho}$ can be used to visualize how it is distributed in the phase space. Thus, in order to find the eigenstates scarred by those families of POs, we make use of the (unnormalized) Husimi function of a state $\hat{\rho}$,
\begin{equation}
\mathcal{Q}_{\hat{\rho}}(\bm x) = \bra{\bm x} \hat{\rho} \ket{\bm x}, 
\end{equation}
where $\ket{\bm x}$ is the coherent state centered at $\bm x$. In the case of a pure state $\hat{\rho}=\dyad{\psi}$, the function reduces to 
\begin{equation}
\mathcal{Q}_{\psi}(\bm x) = \abs{\braket{\psi}{\bm x}}^2.
\label{eq:Hu_pure_state}
\end{equation}

To quantify the degree of scarring of a given state by a specific PO, we consider the temporal average of the state's Husimi function along that PO. For a state $\hat{\rho}$ and the PO $\mathcal{O}$ with period $T$, we define the quantity
\begin{equation}
\left\langle\mathcal{Q}_{\hat{\rho}} \right \rangle_{\mathcal{O}} = \frac{1}{T}\int_0^T \dif  t \, \mathcal{Q}_{\hat{\rho}}(\bm x(t)),
\label{eqn:orbpop}
\end{equation}
where $\bm x(t) \in \mathcal{O}$ and any initial point $\bm x(0) \in \mathcal{O}$ can be used to perform the average. This  measure is similar to the tube phase-space projection introduced in Ref.~\cite{Kaplan1998}, but now taking into account the temporal behavior of the orbit. From the definition, we have that
\begin{equation}
\left\langle\mathcal{Q}_{\hat{\rho}}\right \rangle_{\mathcal{O}} = \tr(\hat{\rho}\,\hat{\rho}_\mathcal{O}),
\label{eq:orbitpopulationasoverlap}
\end{equation}
where 
\begin{equation}
\label{eqn:rho_O}
\hat{\rho}_\mathcal{O}= \frac{1}{T}\int_0^T \dif  t \, \dyad{\bm x(t)}
\end{equation}
is a tubular Gaussian distribution around the PO and $\ket{\bm x(t)}$ is the coherent state centered at the point $\bm x(t)$. 
See App.~\ref{app:ScarringAsOverlap} for an illustration of the Husimi function of $\hat{\rho}_\mathcal{O}$.

From Eq.~\eqref{eq:orbitpopulationasoverlap}, we can see that $\left\langle\mathcal{Q}_{\hat{\rho}} \right \rangle_{\mathcal{O}}$ is the overlap of state $\hat{\rho}$ with $\hat{\rho}_\mathcal{O}$. To construct a baseline to compare the value of $\left\langle\mathcal{Q}_{\hat{\rho}} \right \rangle_{\mathcal{O}}$ with, we consider a totally delocalized state
\begin{equation}
\hat{\rho}_\epsilon = \frac{1}{\mathcal{V}(\epsilon)}\int_\mathcal{M} \dif \bm{x} \dyad{\bm x}\delta(\epsilon-h_\text{cl}(\bm x))
\end{equation}
comprised of all the coherent states within a single energy shell at $\epsilon=h_\text{cl}(\mathcal{O})$, where $\mathcal{V}(\epsilon)=\int_\mathcal{M}\dif \bm x \,\delta( \epsilon-h_\text{cl}(\bm x))$ is the phase-space volume of the energy shell at $\epsilon$. The value of the trace $\tr(\hat{\rho}_\epsilon \, \hat{\rho}_{\mathcal{O}})$ gives the overlap between a totally delocalized state and the orbit. By defining
\begin{equation}
\mathcal{P}(\mathcal{O},\hat{\rho})=\frac{\tr(\hat{\rho} \, \hat{\rho}_{\mathcal{O}})}{\tr(\hat{\rho}_\epsilon  \hat{\rho}_{\mathcal{O}})} ,
\label{eqn:P}
\end{equation}
we obtain a direct measure of quantum scarring. A value of $\mathcal{P}(\mathcal{O},\hat{\rho})=1$ indicates that the overlap between state $\hat{\rho}$ and the PO $\mathcal{O}$ is equal to that of a totally delocalized state.  Values greater than $1$ indicate that the state $\hat{\rho}$ is scarred by the PO $\mathcal{O}$. A value of $\mathcal{P}(\mathcal{O},\hat{\rho})=2$, for example, says that state $\hat{\rho}$ is twice as likely to be found near the PO as compared to the delocalized state $\hat{\rho}_\epsilon$.  Values less than $1$ signal that state $\hat{\rho}$ is less likely to be found near the PO than a fully delocalized state. This avoidance of a specific PO may be regarded as an ``anti-scarring'' effect.

We may quantify how much a state $\hat{\rho}$ is scarred by the POs of the families described in Sec.~\ref{sec:UPOS} by defining the numbers
\begin{align}
\label{eq:PABDefinition}
\mathcal{P}^{A}(\epsilon, \hat{\rho})=\mathcal{P}(\mathcal{O}_{\epsilon}^A,\hat{\rho}) +\mathcal{P}(\widetilde{\mathcal{O}}_{\epsilon}^A,\hat{\rho}) &&\text{and}&&
\mathcal{P}^{B}(\epsilon, \hat{\rho})=\mathcal{P}(\mathcal{O}_{\epsilon}^B, \hat{\rho}) +\mathcal{P}(\widetilde{\mathcal{O}}_{\epsilon}^B,\hat{\rho}) 
\end{align}
for any energy $\epsilon$. In words, $\mathcal{P}^{A}$ measures how much $\hat{\rho}$ populates the region of the phase space visited by the POs of families $\mathcal{A}$ and $\widetilde{\mathcal{A}}$ at the energy  $\epsilon$, while $\mathcal{P}^{B}$ does the same for the families $\mathcal{B}$ and $\widetilde{\mathcal{B}}$.

\subsubsection{Overlap of the energy eigenstates with classical periodic orbits}
\label{subsub}

\begin{figure*}[ht]
\centering
\begin{tabular}{l}
\includegraphics[width=1\textwidth]{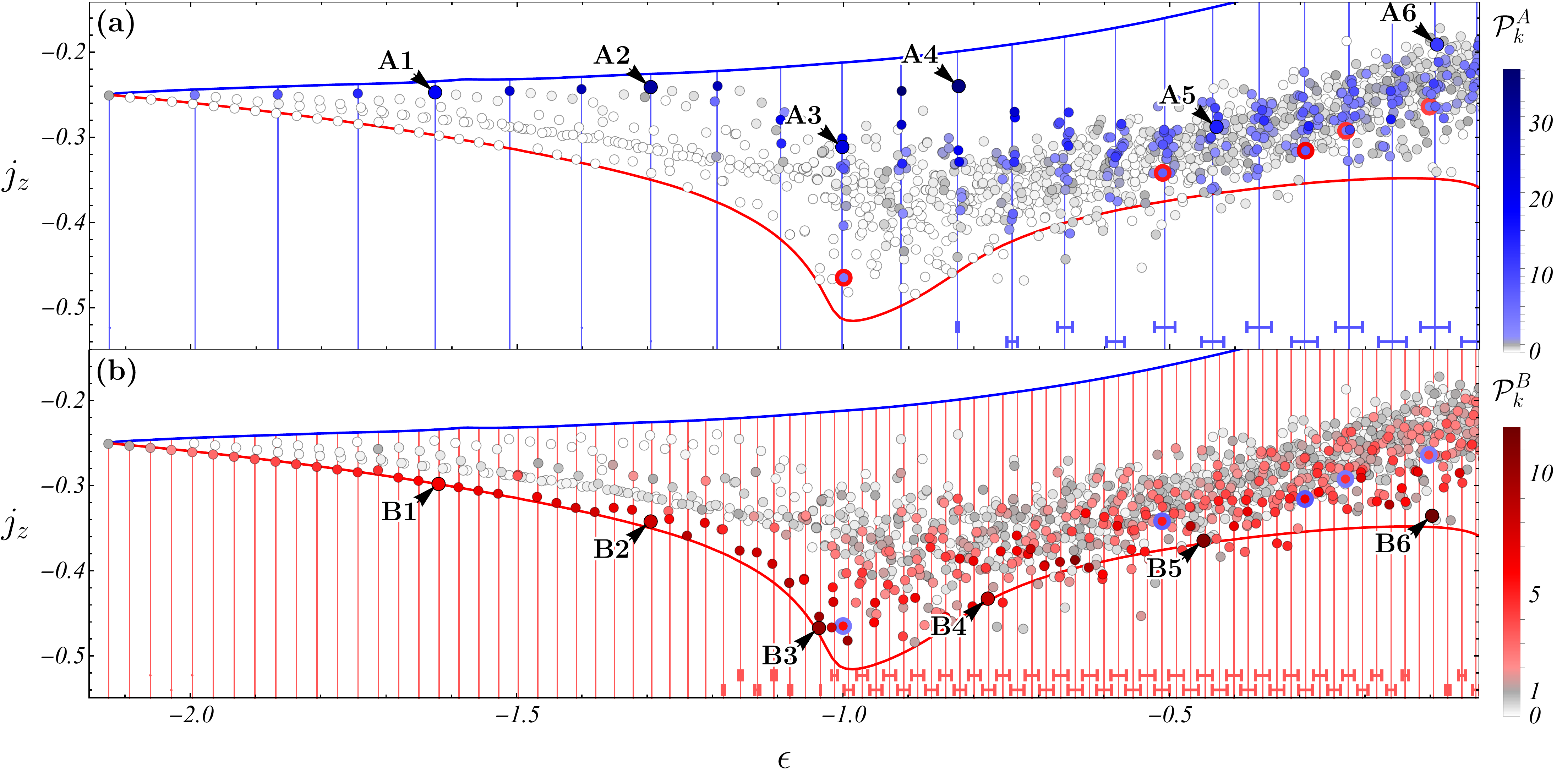}\\
\includegraphics[width=1\textwidth]{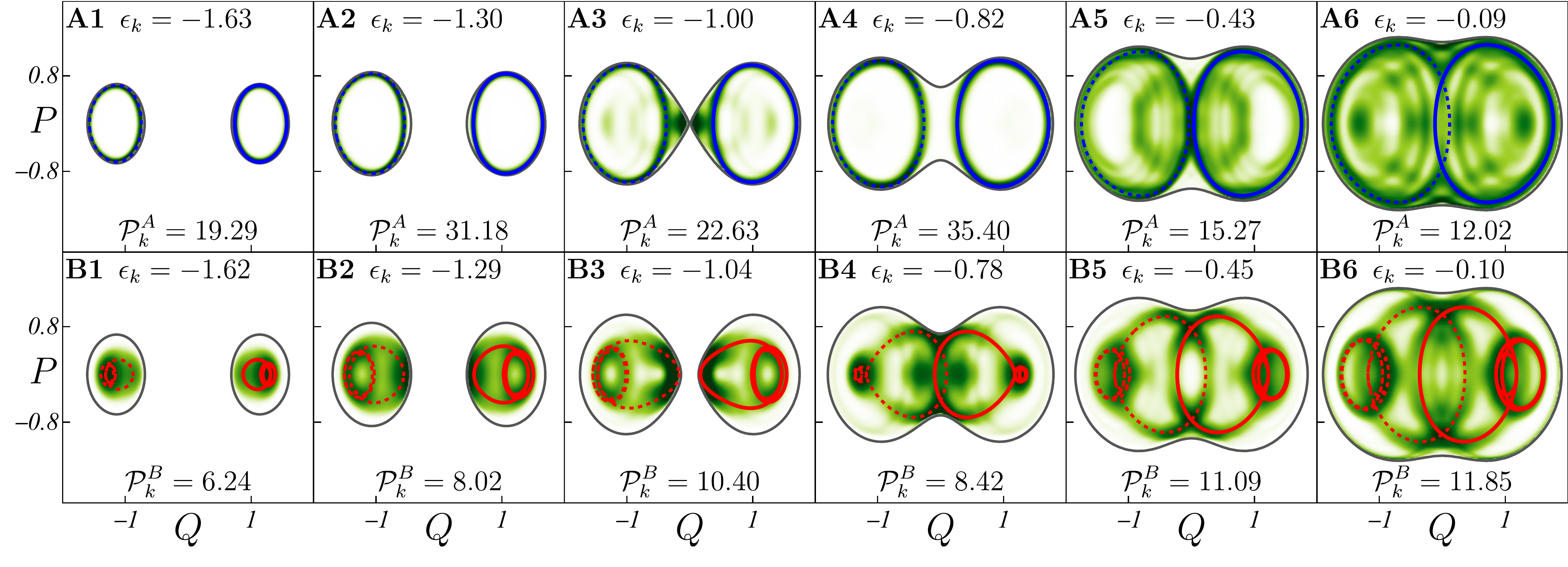} 

\end{tabular}
\caption{Panels (a) and (b): The expectation values of the operator $\hat{j}_z=\hat{J}_z/j$ for each energy eigenstate $\ket{E_k}$ are indicated with circles. Their colors correspond to the value of $\mathcal{P}^A_{k}$ (a) (blue) and $\mathcal{P}^B_{k}$ (b) (red). For panel (a) [(b)], white corresponds to value 0, grey to 1, and larger values are blue [red]. The five eigenstates marked with circles of both colors, that is blue circles outlined by red in (a) [red circles outlined by blue in (b)] are the ones where $\mathcal{P}_k^A$ and $\mathcal{P}_k^B$ are both greater than $4$.
Thick solid curves are the classical average of $j_z$ over the POs of the families $\mathcal{A}$ (blue) and  $\mathcal{B}$ (red) given by Eq.~\eqref{eqn:jz_classical_expectation}. 
In (a) [(b)], the vertical thin blue [red] lines represent the semiclassically quantized energies $\mathcal{E}_i^{A}$  [$\mathcal{E}_i^{B}$] obeying Eq.~\eqref{eqn:BohrSommerfeldPeriod}.  
The small horizontal bars at the bottom of panels (a) and (b) have width $2\lambda^A_{\epsilon_i^{A}}/j$ (blue) and $2\lambda^B_{\epsilon_i^{B}}/j$ (red), where $\lambda^{A,B}_{\epsilon}$ is the Lyapunov exponent of the PO $\mathcal{O}^{A,B}_{\epsilon}$.  
\\Bottom panels A1-B6: The green shades indicate the projected Husimi distribution $\widetilde{\mathcal{Q}}_{E_k}(Q,P)$ for the corresponding eigenstates marked in panels (a) and (b) (darker colors indicate higher values). The gray curved outlines mark the border of the energy shell. The colored lines in the panels A1-B6 draw the POs: $\mathcal{O}^A_{\epsilon_k}$ (solid blue), $\mathcal{O}^B_{\epsilon_k}$ (solid red), $\widetilde{\mathcal{O}}^A_{\epsilon_k}$ (dashed blue), $\widetilde{\mathcal{O}}^B_{\epsilon_k}$ (dashed red). The value of $\epsilon_k=E_k/j$ and $\mathcal{P}_k^{A,B}$ for the corresponding eigenstate are shown in each panel. }
\label{fig:02}
\end{figure*}


For an eigenstate $\hat{\rho}_k=\dyad{E_k}$ of the Dicke Hamiltonian $\hat{H}_D$ with scaled eigenenergy $\epsilon_k=E_k/j$, the numbers
\begin{align}
\label{eq:PkDefinition}
&&\mathcal{P}^{A}_{k}=\mathcal{P}^A(\epsilon_k,\hat{\rho}_k) &&\text{and}&&\mathcal{P}^{B}_{k}=\mathcal{P}^B(\epsilon_k, \hat{\rho}_k)&&
\end{align}
measure the scarring produced by the orbits of families $(\mathcal{A},\widetilde{\mathcal{A}})$ and $(\mathcal{B},\widetilde{\mathcal{B}})$, respectively \cite{footNote}.

In Fig.~\ref{fig:02}, we show the expectation value of the operator $\hat{j}_z=\hat{J}_z/j$ for each energy eigenstate, leading to an arrangement known as a Peres lattice~\cite{Peres1984PRL,Bastarrachea2014b}. This is a convenient way to get information about all the eigenstates in a single picture. In the low-energy regime the points are clearly separated and arranged in a lattice-like fashion, but as the energy increases the structure gets destroyed and the number of points becomes much denser, as typical of chaotic systems. In Fig.~\ref{fig:02}~(a), we color the points according to the degree of scarring of each eigenstate $\ket{E_k}$ by the family $\mathcal{A}$, that is, according to the value of  $\mathcal{P}^{A}_{k}$. The same is done in Fig.~\ref{fig:02}~(b) for family $\mathcal{B}$ using now $\mathcal{P}^{B}_{k}$. 

As the phase-space distribution of a scarred state is close to a PO, the expectation value of the observable $\hat{j}_z$ should be similar to the average of the classical variable $j_z= (Q^2 + P^2)/2$ over the PO. In Fig.~\ref{fig:02}~(a) and Fig.~\ref{fig:02}~(b), we plot with blue and red thick lines the value of the classical average
\begin{equation}
\left\langle j_z \right\rangle_{\mathcal{O}^{A,B}_\epsilon} = \frac{1}{T^{A,B}_\epsilon}\int_0^{T^{A,B}_\epsilon} \dif t\, j_z(\bm x(t)),
\label{eqn:jz_classical_expectation}
\end{equation} 
over each PO from family $\mathcal{A}$ (blue) and $\mathcal{B}$ (red),
where $T^{A,B}_\epsilon$ is the period of $\mathcal{O}^{A,B}_{\epsilon}$. We see in Fig.~\ref{fig:02}~(a) [Fig.~\ref{fig:02}~(b)] that for the low energies in the regular regime, the eigenstates with a high value of $\mathcal{P}^{A}_{k}$ [$\mathcal{P}^{B}_{k}$] lie very close to this blue [red] line. 

Using the Bohr-Sommerfeld quantization rule, we semiclassically quantize the energies of the families $\mathcal{A}$ and $\mathcal{B}$. Beginning from the ground-state energy $\mathcal{E}_0^A=\mathcal{E}_0^B=\epsilon_{\text{GS}}$,
we successively find the excited energies $\mathcal{E}_{i}^{A,B}$ by applying the quantization condition, 
\begin{equation}
\int_{\mathcal{E}_{i-1}^{A,B}}^{\mathcal{E}_{i}^{A,B}}\dif \epsilon \, T^{A,B}_\epsilon =2\pi \hbar_{\text{eff}}=\frac{2\pi}{j}.
\label{eqn:BohrSommerfeldPeriod}
\end{equation}
We plot the obtained energies $\mathcal{E}_i^{A}$ [$\mathcal{E}_i^{B}$] with vertical thin blue [red] lines in Fig.~\ref{fig:02}~(a) [Fig.~\ref{fig:02}~(b)]. These vertical lines coincide perfectly with the individual eigenstates scarred by the POs of the families $\mathcal{A}$ and $\mathcal{B}$ in the low-energy region, where the POs are stable. As the energy increases and the POs become unstable, we find clusters of scarred eigenstates distributed around the semiclassical energies. According to Gutzwiller trace formula for the density of states \cite{Gutzwiller1971} and also to Ref.~\cite{Heller1991}, 
the width of these clusters should be given by the Lyapunov exponent of the PO at the respective energy, which is indeed confirmed by the horizontal bars shown at the bottom of Fig.~\ref{fig:02}~(a) and Fig.~\ref{fig:02}~(b), whose width is twice the value of the Lyapunov exponents of $\mathcal{O}^{A}_{\mathcal{E}_i^{A}}$ (blue) and $\mathcal{O}^{B}_{\mathcal{E}_i^{B}}$ (red) multiplied by $\hbar_\text{eff}=1/j$.

The distribution of scarred states in the spectrum is governed by two numbers: (i) the periods of the POs determine the semiclassically quantized energies $\mathcal{E}_i$ around which clusters of scarred states appear, and (ii) the Lyapunov exponents control the width of these clusters. These two numbers behave differently for families $\mathcal{A}$ and $\mathcal{B}$ as explained below.
\setlength{\leftmargini}{15pt}
\begin{enumerate}
\item[(i)] The periods of the POs in family $\mathcal{A}$ are between three and four times shorter than those of $\mathcal{B}$. This has two consequences, $\mathcal{P}_k^A$ is higher than $\mathcal{P}_k^B$ and there is a smaller number of eigenstates scarred by family $\mathcal{A}$ as compared to $\mathcal{B}$. The former occurs because the POs from family $\mathcal{A}$ are shorter in the phase space than those from family $\mathcal{B}$, thus having smaller overlaps with a delocalized state. Since the denominator in $\mathcal{P}(\mathcal{O},\hat{\rho})$ in Eq.~(\ref{eqn:P}) is the overlap of the PO with a delocalized state, this results in the values of $\mathcal{P}^{A}_{k}$ being between three to four times higher than $\mathcal{P}^{B}_{k}$. The latter occurs because a shorter period translates in greater energy separations $\mathcal{E}_{i+1}-\mathcal{E}_i$. Thus, there are approximately four times more red vertical lines in Fig. \ref{fig:02}~(b), marking $\mathcal{E}_i^B$, than blue vertical lines in Fig. \ref{fig:02}~(a), marking $\mathcal{E}_i^A$. Because scarred eigenstates appear close to these energies, this results in more states scarred by family $\mathcal{B}$ than states scarred by family $\mathcal{A}$.

\item[(ii)] The Lyapunov times of the POs in family $\mathcal{A}$ are always much larger than their periods, which reflects the fact that the energy differences $\mathcal{E}_{i+1}^{A}-\mathcal{E}_i^{A}$ are larger than the Lyapunov exponents multiplied by $\hbar_\text{eff}$. For family $\mathcal{B}$, the period of its POs in the chaotic region are only slightly shorter than the respective Lyapunov times. This allows us to clearly identify the clusters of eigenstates scarred by family $\mathcal{A}$ around the semiclassical energies $\mathcal{E}_i^{A}$ in Fig.~\ref{fig:02}~(a), but makes it harder to distinguish the ones scarred by $\mathcal{B}$ around $\mathcal{E}_i^{B}$ in Fig.~\ref{fig:02}~(b), because neighboring clusters may overlap. 
\end{enumerate}

We close Sec.~\ref{subsub} with the interesting observation that POs from both families may affect the same eigenstate. In Fig.~\ref{fig:02}~(a) [Fig.~\ref{fig:02}~(b)], the blue circles outlined by red [red circles outlined by blue] mark the eigenstates where both $\mathcal{P}_k^A$ and $\mathcal{P}_k^B$ are greater than $4$. These five states are located at energies where the semiclassical quantizations of both families coincide.

\subsubsection{Projected Husimi distribution}

We now use the Husimi distribution to visually confirm that the energy eigenstates with large values of $\mathcal{P}^{A}_{k}$ and $\mathcal{P}^{B}_{k}$ are scarred by the classical POs of families $\mathcal{A}$ and $\mathcal{B}$. 

For a state $\hat{\rho}$ and energy $\epsilon$, we consider the projection of the Husimi function $\mathcal{Q}_{\hat{\rho}}$ over the classical energy shell $h_\text{cl}(\bm{x})=\epsilon$ by integrating out the bosonic variables $(q,p)$,
\begin{equation}
\label{eqn:ProjHusimiDef}
\widetilde{\mathcal{Q}}_{\epsilon,\hat{\rho}}(Q,P) = \iint \dif q \dif p \, \delta \big(\epsilon - h_\text{cl}(\bm{x}) \big) \mathcal{Q}_{\hat{\rho}}(\bm{x}),
\end{equation}
where $\bm{x}=(q,p;Q,P)$.
For an eigenstate $\hat{\rho}_k=\dyad{E_k}$,  the projection $\widetilde{\mathcal{Q}}_k=\widetilde{\mathcal{Q}}_{\epsilon_k,\hat{\rho}_k}$ yields a function depending only on the variables $(Q,P)$, which can be compared with the projection of the POs over the same plane $Q$-$P$ (see App.~\ref{app:ProjHusimi} for details on the computation of this projection). 
 
We select 12 energy eigenstates \cite{PilatowskyARXIV}, marked as A1-A6 and B1-B6 in Fig.~\ref{fig:02}~(a) and Figs.~\ref{fig:02}~(b), which in addition to having high values of $\mathcal{P}^{A,B}_{k}$ lie close to the classical average of $j_z$ given by Eq.~\eqref{eqn:jz_classical_expectation}. We plot the Husimi projection $\widetilde{\mathcal{Q}}_{k}(Q,P)$ for each one of these eigenstates at the 12 bottom panels of Fig.~\ref{fig:02}. These distributions are superposed by the projections of the  POs $\mathcal{O}^A_{\epsilon_k}$ (blue solid line), $\mathcal{O}^B_{\epsilon_k}$ (red solid line), $\widetilde{\mathcal{O}}^A_{\epsilon_k}$ (blue dashed line), and $\widetilde{\mathcal{O}}^B_{\epsilon_k}$ (red dashed line). Scarring  is clearly visible in all panels. The quantum states A1-A6 and B1-B6 are highly concentrated around the classical periodic orbits. This happens even in the chaotic region of high excitation energy, as seen for A5, A6, B5, and B6 with $\epsilon_k>-0.5$, where the classical dynamics is ergodic

Interesting features are revealed by the juxtaposition of the Husimi projections and the periodic orbits. For example, the eigenstate B3 shows a significant concentration of probability towards the center at $Q=P=0$. This is because close to the energy of this state, specifically at $q=p=Q=P=0$ and $\epsilon=-1$, there is an unstable stationary point and so the dynamics of the periodic orbit slows down around it. This gets reflected in the eigenstate by the localization of the Husimi distribution in the same region [see Fig.~\ref{fig:06}~(b) in App.~\ref{app:ScarringAsOverlap} for an illustration of this effect]. Also noticeable is the fact that the probability distributions of the eigenstates A5 and A6 are not entirely confined to the periodic orbit, but extend beyond it, which contrasts with the high density concentration of the eigenstates A1-A4. This results in lower values of $\mathcal{P}^A_k$ for A5 and A6 as compared to A1-A4.


\section{SCARRING AND DYNAMICS}
\label{sec:dynamics}

In this section, we show the effects that the scarred states have over the dynamical properties of non-stationary states. We consider three initial Glauber-Bloch coherent states that have energy in the chaotic regime. One is centered in a point of a UPO of family $\mathcal{A}$, one in a point of a UPO of family $\mathcal{B}$, and the third one is away from the POs of the identified families $\mathcal{A}$, $\mathcal{B}$, $\widetilde{\mathcal{A}}$, and $\widetilde{\mathcal{B}}$ .


\begin{figure*}[ht]
\centering
\includegraphics[width=1\textwidth]{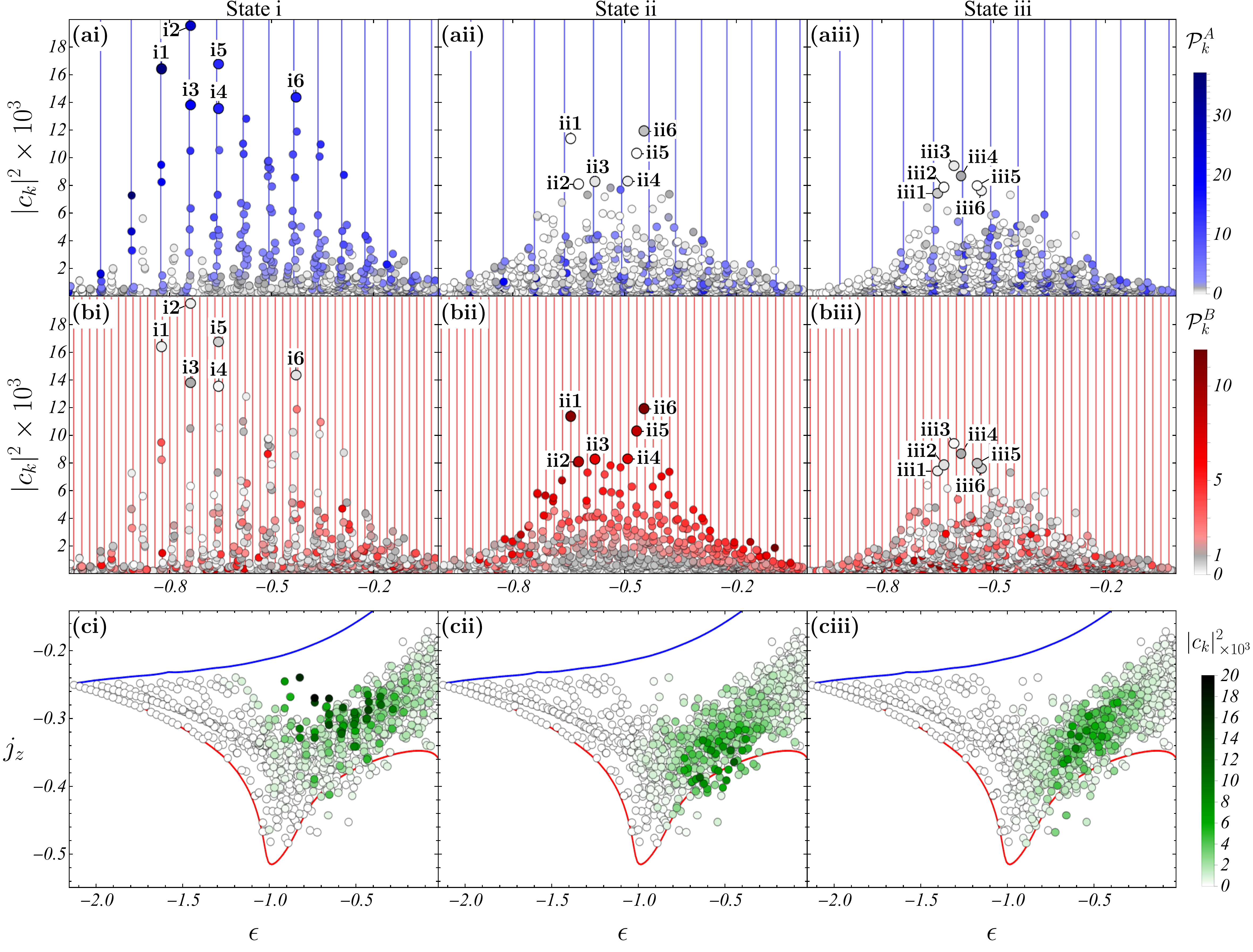}
\caption{Panels (ai)-(biii): LDoS for coherent states $\ket{\bm x_\text{i}}$ (ai and bi), $\ket{\bm x_\text{ii}}$ (aii and bii), and $\ket{\bm x_\text{iii}}$ (aiii and biii). The circles indicating the eigenstates are colored by the value of $\mathcal{P}^A_k$ (ai, aii, aiii) and $\mathcal{P}^B_k$ (bi, bii, biii). The vertical lines show the semiclassically quantized energies $\mathcal{E}^A_i$ in (ai)-(aiii) and $\mathcal{E}^B_i$ in (bi)-(biii). The indicated eigenstates i1-i6, ii1-ii6 and iii1-iii6 are the ones with the highest participation in the LDoS of the corresponding coherent state. 
Panels (ci)-(ciii): Peres lattice for the operator $\hat{j}_z$. The tones of green indicate the values of $\abs{c_k}^2$ for the LDoS of coherent states $\ket{\bm x_\text{i}}$ (ci), $\ket{\bm x_\text{ii}}$ (cii), and $\ket{\bm x_\text{iii}}$ (ciii). The solid lines correspond to the classical average of $j_z$ over the POs of the families $\mathcal{A}$ (blue) and $\mathcal{B}$ (red), as given by Eq.~\eqref{eqn:jz_classical_expectation}.}
\label{fig:03}
\end{figure*}


\subsection{Initial coherent states}
 
We  expand the selected coherent states in the Hamiltonian  eigenbasis 
\begin{equation}
\label{eqn:EigenCoh}
\ket{\bm x}=\sum_k c_k \ket{E_k}.
\end{equation}
The amplitudes $c_k$ and the degree of scarring of the corresponding eigenstates $\ket{E_k}$ determine the dynamical properties of the initial  states.

We consider three initial coherent states, $\ket{\bm x_\text{i}}$, $\ket{\bm x_\text{ii}}$, and $\ket{\bm x_\text{iii}}$, where 
\begin{align}
\label{eqn:CohSta}
\bm{x}_{\text{i}} & = \big(q=-0.795,p=0;Q=1.75,\,P=0 \,\big), \\
\bm{x}_{\text{ii}} & = \big(q=-0.105,p=0;Q=0.9,P=0.7\big), \nonumber \\
\bm{x}_{\text{iii}} & = \big(q=0.624,\,\,\,\,\,p=0;Q=-0.2,P=1\big).\nonumber
\end{align}
They all have mean energy in the chaotic region, $\langle \hat{H}_D \rangle/j=\epsilon_{\text{i}}=\epsilon_{\text{ii}}=\epsilon_{\text{iii}}=-0.5$, and the widths of their energy distributions in the energy eigenbasis are $(\sigma_{\text{i}},\sigma_{\text{ii}},\sigma_{\text{iii}})=(0.248,0.212,0.192)$ (in units of $\epsilon$)~\cite{Schliemann2015,Lerma2018}.
The points $\bm{x}_{\text{i}}$, $\bm{x}_{\text{ii}}$, and $\bm{x}_{\text{iii}}$ are marked with little circles in Figs.~\ref{fig:01}~(a1), (a2), (b1), and (b2). They all have the same shade of green indicating their equal energy. The point $\bm x_\text{i}$ sits on top of the PO of family $\mathcal{A}$ that has that same energy $-0.5$, as can be seen in Figs.~\ref{fig:01}~(a1) and (a2). The point $\bm x_\text{ii}$ sits on top of the PO of $\mathcal{B}$, as shown in Figs.~\ref{fig:01}~(b1) and (b2). The point $\bm{x}_{\text{iii}}$ is located far away from the POs of both families (and the mirrored families), as seen in Figs.~\ref{fig:01}~(a1), (a2), (b1), and (b2).

\subsubsection{Scarring of coherent states}

In Figs.~\ref{fig:03}~(ai)-(aiii) and Figs.~\ref{fig:03}~(bi)-(biii), we plot the energy distribution,

\begin{equation}
\label{eqn:LDoS}
\mathcal{G}(\epsilon)=\sum_{k}\abs{c_{k}}^{2}\delta(\epsilon-\epsilon_{k}),
\end{equation}
of the initial coherent states $\ket{\bm x_\text{i}}$ [Figs.~\ref{fig:03}~(ai) and (bi)], $\ket{\bm x_\text{ii}}$ [Figs.~\ref{fig:03}~(aii) and (bii)], and $\ket{\bm x_\text{iii}}$ [Figs.~\ref{fig:03}~(aiii) and (biii)]. These distributions are usually referred to as local density of states (LDoS). In a sense, Figs.~\ref{fig:03} (ai)-(biii) are similar to Figs.~\ref{fig:02}~(a) and (b), but now instead of $j_z$ in the vertical axis, we have the components of the chosen coherent states. The distributions in Figs.~\ref{fig:03}~(ai)-(aiii) are the same as in Figs.~\ref{fig:03}~(bi)-(biii), what changes is just the colors: the blue tones in Figs.~\ref{fig:03}~(ai)-(aiii) indicate the values of $\mathcal{P}_k^A$ and the red tones in Figs.~\ref{fig:03}~(bi)-(biii) indicate the values of $\mathcal{P}_k^B$ [see Eq.~\eqref{eq:PkDefinition}]. In addition to the circles corresponding to amplitudes of the weights $\abs{c_k}^2$, Figs.~\ref{fig:03}~(ai)-(aiii)  [Figs.~\ref{fig:03}~(bi)-(biii)] also display vertical lines that mark the semiclassically quantized energies $\mathcal{E}_i^A$ [$\mathcal{E}_i^B$] given by Eq.~\eqref{eqn:BohrSommerfeldPeriod}.

We observe the following features for the three initial coherent states:
\setlength{\leftmargini}{20pt}
\begin{enumerate}
\item[(i)] For the initial coherent state $\ket{\bm x_\text{i}}$ located in the PO of family $\mathcal{A}$, the largest components $\abs{c_k}^2$, indicated as i1-i6 in Fig.~\ref{fig:03}~(ai), correspond to the eigenstates with large values of $\mathcal{P}_k^A$ and therefore scarred by the POs of family $\mathcal{A}$. Due to the high participation of these eigenstates, we say that the initial coherent state $\ket{\bm x_\text{i}}$ is itself scarred by family $\mathcal{A}$. As visible in Fig.~\ref{fig:03}~(ai), the LDoS of $\ket{\bm x_\text{i}}$ exhibits a clear comb-like pattern, which is typical of scarred states \cite{Heller1991}. As seen in Fig. \ref{fig:02} (a), the scarred eigenstates cluster around the semiclassical energies $\mathcal{E}_i^A$. Because these scarred eigenstates give the largest contributions to the initial state, that is they lead to the biggest components $\abs{c_k}^2$, the LDoS attains higher values around the energies $\mathcal{E}_i^A$, creating the comb-like structure.

One sees that the contributions to $\ket{\bm x_\text{i}}$ from eigenstates non-scarred by the POs of family $\mathcal{A}$ are erratically distributed, with small and medium values of $\abs{c_k}^2$, as evident from the red points in Fig.~\ref{fig:03}~(bi), which mark the eigenstates according to their values of $\mathcal{P}_k^B$. 

\item[(ii)] A similar picture emerges for the initial coherent state $\ket{\bm x_\text{ii}}$ located in the PO of family $\mathcal{B}$. Its largest components $\abs{c_k}^2$, indicated as ii1-ii6 in Fig.~\ref{fig:03}~(bii), correspond to the eigenstates with large values of $\mathcal{P}_k^B$ and thus scarred by the POs of family $\mathcal{B}$. The contributions from eigenstates non-scarred by this family are smaller and their values fluctuate randomly. The initial coherent state $\ket{\bm x_\text{ii}}$ is therefore scarred by family $\mathcal{B}$. Its comb-like structure is somewhat visible, but, because the separations $\mathcal{E}_i^B-\mathcal{E}_{i+1}^B$ are of the order of the Lyapunov exponents of the POs in family $\mathcal{B}$, it is harder to distinguish it. 

As discussed in the previous section, there are more eigenstates scarred by family $\mathcal{B}$ than by family $\mathcal{A}$, due to the difference in the periods of the POs between the two families. This difference between the two families is evident if we compare the number of blue circles in Fig.~\ref{fig:03}~(ai) with the number of red circles in Fig.~\ref{fig:03}~(bii). The larger number of contributing states from family $\mathcal{B}$ explains why the biggest components in Fig.~\ref{fig:03}~(bii) are smaller than the ones in Fig.~\ref{fig:03}~(ai). 

\item[(iii)] For the initial coherent state $\ket{\bm x_\text{iii}}$, which is located far away from the POs of families $\mathcal{A}$, $\mathcal{B}$, $\widetilde{\mathcal{A}}$ and $\widetilde{\mathcal{B}}$, the largest components $\abs{c_k}^2$, indicated as iii1-iii6 in Figs.~\ref{fig:03}~(aiii) and (biii), are not colored by any of the two families, as expected. 
\end{enumerate}

In Figs.~\ref{fig:03}~(ci)-(ciii), we plot the same Peres lattice of the pseudo-spin operator $\hat{j}_{z}$ that we showed in Fig.~\ref{fig:02}, but we now use tones of green to indicate the values of $\abs{c_k}^2$ for the coherent states $\ket{\bm x_\text{i}}$ [Fig.~\ref{fig:03}~(ci)], $\ket{\bm x_\text{ii}}$ [Fig.~\ref{fig:03}~(cii)], and $\ket{\bm x_\text{iii}}$ [Fig.~\ref{fig:03}~(ciii)]. The panels make it evident that the components of the three initial states, all covering a wide range of energies within the chaotic regime, are actually localized in different regions of the lattice. The components of $\ket{\bm x_\text{i}}$ ($\ket{\bm x_\text{ii}}$) concentrate towards the top (bottom) of the lattice, because this is where the eigenstates scarred by family $\mathcal{A}$ ($\mathcal{B}$) tend to be found, closer to the classical average of $j_z$ (Eq. \eqref{eqn:jz_classical_expectation}) over the corresponding family [blue (red) line in Figs.~\ref{fig:03}~(ci)-(ciii)]. The components of $\ket{\bm x_\text{iii}}$ spread more homogeneously over the middle part of the lattice, as seen in Fig.~\ref{fig:03}~(ciii). 


\begin{figure*}[ht]
\centering
\includegraphics[width=1\textwidth]{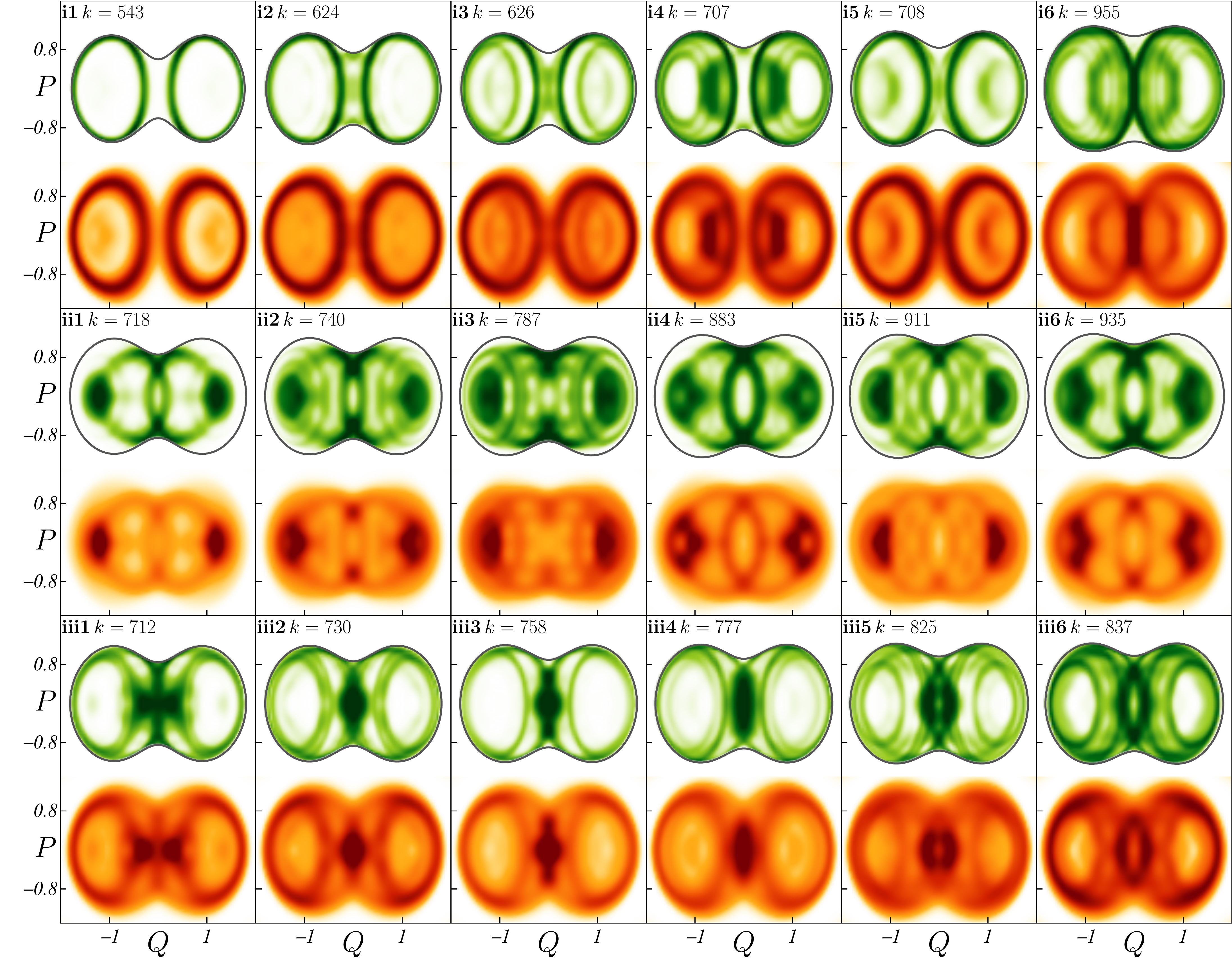}
\caption{Top green plots (bottom orange plots): Projected Husimi distribution $\widetilde{\mathcal{Q}}_k(Q,P)$ over the classical energy shell of the eigenenergy $h_{\text{cl}}(\bm{x})=\epsilon_{k}$ (over all energy shells $\iint \dif q\dif p \mathcal{Q}_k(\bm x)$ \cite{Deaguiar1991,Furuya1992}) for the eigenstates i1-i6, ii1-ii6 and iii1-iii6, which are marked in Figs.~\ref{fig:03}~(ai)-(biii) and have the highest participation in the LDoS of the corresponding coherent state i (top), ii (middle) and iii (bottom). The gray curves outline the border of the energy shell.
In all panels, darker colors indicate larger probabilities.  The index $k$ of the corresponding eigenstate is shown at the top of each panel.}
\label{fig:04}
\end{figure*}


\subsubsection{Husimi distributions of the most contributing eigenstates}

In Fig.~\ref{fig:04}, we plot the projected Husimi distributions for each of the six eigenstates with the largest components in the LDoS of the initial coherent states $\ket{\bm x_\text{i}}$ [Figs.~\ref{fig:04} (i1)-(i6)], $\ket{\bm x_\text{ii}}$ [Figs.~\ref{fig:04} (ii1)-(ii6)], and $\ket{\bm x_\text{iii}}$ [Figs.~\ref{fig:04} (iii1)-(iii6)]. This is done in two different ways: 

(a) By first intersecting the Husimi distribution of the eigenstate  with the energy shell at the respective eigenenergy $h_{\text{cl}}(\bm{x})=\epsilon_{k}$, and then integrating over ($q$,$p$), as we did in Fig.~\ref{fig:02}~A1-B6 [see Eq.~\eqref{eqn:ProjHusimiDef}] and in \cite{PilatowskyARXIV}. This choice is shown in green in the top of each panel of  Fig.~\ref{fig:04}. 

(b) By directly integrating over the bosonic variables $(q,p)$ of the Husimi function, $\iint \dif q\dif p \mathcal{Q}_k(\bm x)$, as done in \cite{Deaguiar1991,Furuya1992, Wang2020}. This alternative is plotted in orange in the bottom of each panel of Fig.~\ref{fig:04}.

As expected, the Husimi functions of the eigenstates i1-i6 [ii1-ii6] with the largest components in the coherent state $\ket{\bm x_\text{i}}$ [$\ket{\bm x_\text{ii}}$] concentrate along the POs of family $\mathcal{A}$ [$\mathcal{B}$]  at the corresponding energy.  This is more evident with the Husimi functions intersected by the energy shells (top green plots in Fig. \ref{fig:04}) than in the complete projected Husimi functions (bottom orange plots in Fig. \ref{fig:04}), which are more blurred, because they include all energy shells. This difference shows the advantage of the projection method (a), which makes it easier to distinguish the outline of the POs. Method (a) was first developed in \cite{PilatowskyARXIV} and more details are given in App.~\ref{app:ProjHusimi}.

Similarly to the eigenstates i1-i6  and ii1-ii6,  the Husimi functions of the eigenstates iii1-iii6 that contribute the most to the initial state $\ket{\bm x_\text{iii}}$ are not smoothly distributed either. As seen in Figs.~\ref{fig:04}~(iii1)-(iii2), the visible concentrations in parts of the phase space suggest that these eigenstates are also scarred by one or more families of POs , although they are different from the families $\mathcal{A}$ and  $\mathcal{B}$.

\subsection{Survival probability of coherent states}

To study the dynamics of the three selected initial states, we consider the survival probability defined as
\begin{equation}
S_{P}(t)=|\langle\bm{x}|\hat{U}(t)|\bm{x}\rangle|^{2},
\end{equation}
where $\hat{U}(t)=e^{-i\hat{H}_{D}t}$ is the unitary evolution operator. $S_{P}(t)$ measures the probability of the evolved state $\hat{U}(t)\ket{\bm{x}}$ to return to its initial state $\ket{\bm{x}}$ for any given time. By expanding the coherent state  in the Hamiltonian eigenbasis as in Eq.~\eqref{eqn:EigenCoh}, we can rewrite the survival probability as
\begin{equation}
\label{eqn:NumSP}
S_{P}(t)=\abs{\sum_{k}\abs{c_{k}}^{2}e^{{-i\epsilon_{k}t}/\hbar_\text{eff}}}^{2},
\end{equation}
which can also be written in an integral form,
\begin{equation}
\label{eqn:SP_as_Fourier_of_LDoS}
S_{P}(t)=\abs{\int \dif\epsilon \, \mathcal{G}(\epsilon)e^{-i\epsilon t/\hbar_\text{eff}}}^{2}=\abs{\mathcal{F}[\mathcal{G}](t)}^2,
\end{equation}
which is the squared norm of the Fourier transform of the LDoS, as defined in Eq.~\eqref{eqn:LDoS}.

In general, the evolution of scarred initial states, where the LDoS is fragmented, produces partial revivals at times before the Lyapunov time~\cite{Heller1984} and the saturation values of the survival probability are larger than those for non-scarred states~\cite{Kaplan1998,Villasenor2020}.  The reason for this behavior is, of course, the low number of eigenstates that contribute to the dynamics. 

 
\begin{figure*}[ht]
\centering
\includegraphics[width=1\textwidth]{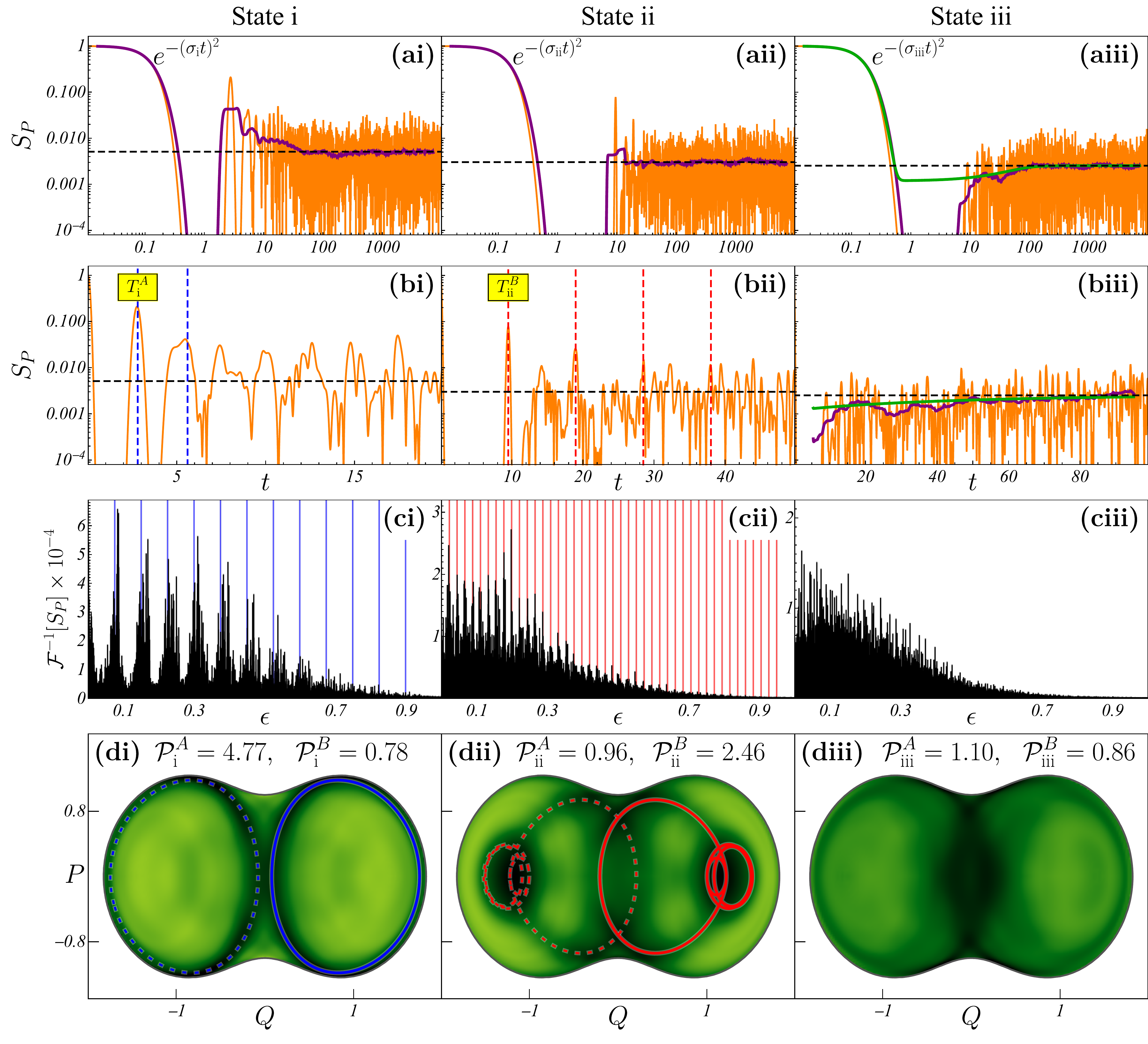}
\caption{Survival probabilities for coherent states $\ket{\bm x_\text{i}}$ [(ai)-(bi)], $\ket{\bm x_\text{ii}}$ [(aii)-(bii)], and $\ket{\bm x_\text{iii}}$ [(aiii)-(biii)] (orange solid curves). Panels (ai)-(aiii) use a log-log scale and (bi)-(biii) a lin-log scale. The thicker purple curves in (ai), (aii), (aiii) and (biii) are running averages. The equilibration value is shown as a horizontal black dashed line in all panels. The average period of the POs in family $\mathcal{A}$ ($\mathcal{B}$) over the energy span of state $\ket{\bm x_\text{i}}$ (state $\ket{\bm x_\text{ii}}$), $T^A_\text{i}$ ($T^B_\text{ii}$) and two (four) integer multiples are drawn as vertical blue (red) dashed lines in panel (bi) (panel (bii)). In panels (aiii) and (biii), the thick green curve represents the analytical expression obtained for the evolution of a random state with a Gaussian energy profile~\cite{Villasenor2020}. 
In (ci)-(ciii), the black bars plot the inverse Fourier transform of the survival probability of the corresponding coherent states as a function of $\epsilon$. The vertical blue [red] lines in panel (ci) [(cii)] mark integer multiples of $2\pi\hbar_\text{eff}/T_\text{i}^A$ [$2\pi\hbar_\text{eff}/T_\text{ii}^B$].
Panels (di)-(diii) plot the projected Husimi distribution $\widetilde{\mathcal{Q}}_{\epsilon,\overline{\rho}}$ of the infinite-time averages $\overline{\rho}=\overline{\rho}_{\bm_\text{i}}$ (di), $\overline{\rho}_{\bm_\text{ii}}$ (dii), and  $\overline{\rho}_{\bm_\text{ii}}$ (diii) at energy $\epsilon=-0.5$. White (not reached) is zero and darker colors indicate higher concentrations. In (di) we plot the POs  $\mathcal{O}^A_{\epsilon}$ (solid blue) and $\widetilde{\mathcal{O}}^A_{\epsilon}$  (dashed blue), and in (dii) the POs  $\mathcal{O}^B_{\epsilon}$ (solid red) and $\widetilde{\mathcal{O}}^B_{\epsilon}$  (dashed red) ($\epsilon=-0.5$). The values of $\mathcal{P}^{A,B}_\text{i}$ (di), $\mathcal{P}^{A,B}_\text{ii}$ (dii), and $\mathcal{P}^{A,B}_\text{iii}$ (diii) are also shown.}
\label{fig:05}
\end{figure*}


The survival probability of the initial coherent states $\ket{\bm x_\text{i}}$, $\ket{\bm x_\text{i}}$, and $\ket{\bm x_\text{i}}$ are respectively shown in Figs.~\ref{fig:05}~(ai)-(bi), Figs.~\ref{fig:05}~(aii)-(bii), and Figs.~\ref{fig:05}~(aiii)-(biii). The orange curves are numerical results and the thick purple ones are running averages. The initial decay of $S_{P}(t)$ for the three cases is Gaussian, since the envelope of the LDoS is Gaussian, and it then reaches values close to zero, as explained in~\cite{Villasenor2020}. The subsequent behavior differs among the three states.

For state $\ket{\bm x_\text{i}}$, the first revival of the survival probability occurs at $t=2.8$ [Fig.~\ref{fig:05}~(bi)], which is precisely the average of the periods $T_\epsilon^A$ within the energy range of state $\ket{\bm x_\text{i}}$, $T^A_\text{i}$. A second revival is observed  at $t=2 T^A_\text{i}$, while the next ones  do not follow this period, because $T^A_\epsilon$ changes strongly through the energy range of state $\ket{\bm x_\text{i}}$ destroying the periodicity. This may be seen in the left inset of Fig.~\ref{fig:01}~(c), where the little blue square marks the value of $T^A_\text{i}$. For longer times, the survival probability equilibrates, fluctuating around its  asymptotic value drawn with a horizontal black dashed line  \cite{Villasenor2020}.

The first revival of $S_{P}(t)$ for state $\ket{\bm x_\text{ii}}$ appears  at $t = T^B_\text{ii}=9.5$ [Fig.~\ref{fig:05}~(bii)], which corresponds to the average period of the POs in the family $\mathcal{B}$ within the energy range of state $\ket{\bm x_\text{ii}}$. The subsequent revivals happens at multiples of this period. The periods $T^B_\epsilon$ do not vary as much as $T^A_\epsilon$ around $\epsilon=-0.5$, as seen in the right inset of Fig.~\ref{fig:01}~(c), where the value $T^B_\text{ii}$ is marked with a little red square. The slope  of $T^A_\epsilon$ around  $T^A_\text{i}$ is large, while $T^B_\epsilon$ actually attains a local minimum close to $T^B_\text{ii}$. As a result, the revivals of the survival probability of state $\ket{\bm x_\text{ii}}$ follow the period $T^B_\text{ii}$ for much longer than in the case of state $\ket{\bm x_\text{i}}$. 

A great advantage of having identified the families $\mathcal{A}$ and $\mathcal{B}$ is that we know why and where the revivals should happen. Having access to the number of contributing eigenstates and knowing why there are more contributing states from family $\mathcal{B}$ than from family $\mathcal{A}$ help us understand also why the survival probability for the initial state $\ket{\bm x_\text{ii}}$ saturates at a lower point than $S_{P}(t)$ for the state $\ket{\bm x_\text{i}}$. 

In contrast to states $\ket{\bm x_\text{i}}$ and  $\ket{\bm x_\text{ii}}$, the initial coherent state  $\ket{\bm x_\text{iii}}$ does not show revivals. Instead, the survival probability shows a behavior similar to what we obtain when considering an initial state where the coefficients $c_k$ are random numbers. This latter case is indicated with a green line in Fig.~\ref{fig:05}~(aiii) and Fig.~\ref{fig:05}~(biii) and it is well described using random matrix theory~\cite{Lerma2019,Villasenor2020}. 

A complete picture of the frequencies dominating the dynamical behavior of the coherent states may be obtained by calculating the inverse Fourier transform of the survival probability,
\begin{equation}
\label{eqn:inverseFourierofSP}
\mathcal{F}^{-1}[S_P](\epsilon)=\frac{1}{2\pi \hbar_\text{eff}}\int \dif t S_P(t)e^{i \epsilon t / \hbar_\text{eff}}=\int \dif \epsilon' \mathcal{G}(\epsilon')\mathcal{G}(\epsilon' + \epsilon),
\end{equation}
which is the autocorrelation function of the LDoS. This function gives us the distribution of frequencies of the survival probability in units of $\epsilon$. In Figs.~\ref{fig:05}~(ci), (cii), and (ciii), we plot $\mathcal{F}^{-1}[S_P](\epsilon)$ with black bars for states  $\ket{\bm x_\text{i}}$, $\ket{\bm x_\text{ii}}$, and $\ket{\bm x_\text{iii}}$, respectively. The vertical blue [red] lines in  Fig.~\ref{fig:05}~(ci) [Fig.~\ref{fig:05}~(cii)] mark the integer multiples of the energy separation $2\pi\hbar_\text{eff}/T_\text{i}^A$ [$2\pi\hbar_\text{eff}/T_\text{ii}^B$] corresponding to the period $T_\text{i}^A$ [$T_\text{ii}^B$]. In Fig.~\ref{fig:05}~(ci) [Fig.~\ref{fig:05}~(cii)], the function $\mathcal{F}^{-1}[S_P](\epsilon)$ displays a comb-like pattern following the energy separation above, indicating that the period $T_\text{i}^A$ [$T_\text{ii}^B$] indeed  dominates the behavior of the survival probability of state $\ket{\bm x_\text{i}}$ [$\ket{\bm x_\text{ii}}$]. Figure~\ref{fig:05}~(ciii), on the other hand, does not exhibit any recognizable pattern. Because  $\mathcal{F}^{-1}[S_P](\epsilon)$ is the autocorrelation function of the LDoS (Eq.~\ref{eqn:inverseFourierofSP}), it enhances any periodic structure present in $\mathcal{G}(\epsilon)$.  This amplifies the comb-like structures that is present in the LDoS of states $\ket{\bm x_\text{i}}$ and $\ket{\bm x_\text{ii}}$, and confirms that such a structure is absent in the LDoS of state $\ket{\bm x_\text{iii}}$. 

The fact that the survival probability of state $\ket{\bm x_\text{iii}}$ is so well described by random matrix theory and that no frequencies clearly stand out in its Fourier transform is puzzling at first sight, since Figs. \ref{fig:04} (iii1)-(iii6) suggest that the most contributing eigenstates to $\ket{\bm x_\text{iii}}$, that is eigenstates iii1-iii6 in Figs.~\ref{fig:03}~(aiii) and (biii), are also scarred. What we have come to understand from the analysis of various initial states is that the onset of revivals depends on two factors: the eigenstates with the largest participations in the LDoS  should be scarred by the same family of POs, so that the periods of the POs generating the scars are similar; and these periods should be small in comparison to the Lyapunov times. For a given family of POs, if the Lyapunov time is shorter than the period, the revival does not have time to develop before saturation. This happens if either the PO is very unstable or if it has a very long period. It may therefore be that the eigenstates iii1-iii6  either do not belong to the same family and thus the corresponding POs have very dissimilar periods, or that these periods are so long that the revivals are unable to manifest. These are, however, open questions that can only be answered with the identification of the families associated with those states.
\subsection{Dynamical scarring}

We finally analyze how the scarring manifests in the infinite-time average of the coherent states~\cite{PilatowskyARXIV}
\begin{equation}
\overline{\rho}_{\bm x}=\lim_{T\to \infty} \frac{1}{T}\int_0^T \dif t \, \hat{U}(t)\dyad{\bm x}\hat{U}^{\dagger}(t).
\end{equation}

With the functions $\mathcal{P}^A$ and $\mathcal{P}^B$ defined by Eq. \eqref{eq:PABDefinition}, we may calculate
\begin{align*}
\mathcal{P}^A_\text{i}=&\mathcal{P}^A(\epsilon_\text{i}, \overline{\rho}_{\bm x_\text{i}})=4.77&
\mathcal{P}^A_\text{ii}=&\mathcal{P}^A(\epsilon_\text{ii}, \overline{\rho}_{\bm x_\text{ii}})=0.96&\quad
\mathcal{P}^A_\text{iii}=&\mathcal{P}^A(\epsilon_\text{iii}, \overline{\rho}_{\bm x_\text{iii}})=1.10& \\
\mathcal{P}^B_\text{i}=&\mathcal{P}^B(\epsilon_\text{i}, \overline{\rho}_{\bm x_\text{i}})=0.78&
\mathcal{P}^B_\text{ii}=&\mathcal{P}^B(\epsilon_\text{ii}, \overline{\rho}_{\bm x_\text{ii}})=2.46&
\mathcal{P}^B_\text{iii}=&\mathcal{P}^B(\epsilon_\text{iii}, \overline{\rho}_{\bm x_\text{iii}})=0.86.& \nonumber
\end{align*}
Notice that $\mathcal{P}^A_\text{i}=4.77$ and $\mathcal{P}^B_\text{ii}=2.46$ are particularly large, while the rest of the numbers are less or approximately equal to $1$. This means that, at any time $t$, state $\hat{U}(t)\ket{\bm x_\text{i}}$ [$\hat{U}(t)\ket{\bm x_\text{ii}}$] is more likely to be found in the vicinity of the orbits of families $\mathcal{A}$ and $\widetilde{\mathcal{A}}$ [$\mathcal{B}$ and $\widetilde{\mathcal{B}}$] at energy $\epsilon=-0.5$ than what one would expect for a state that is completely delocalized in the energy shell. This effect is known as dynamical scarring \cite{Tomiya2019}.

We can visualize the dynamical scars by calculating the projected Husimi distributions $\mathcal{Q}_{\epsilon,\hat{\rho}}$ [Eq. \eqref{eqn:ProjHusimiDef}] of the infinite-time averages $\hat{\rho}=\overline{\rho}_{\bm x_\text{i}}$, $\overline{\rho}_{\bm x_\text{ii}} $, and $\overline{\rho}_{\bm x_\text{iii}}$. These are plotted in Figs. \ref{fig:05} (di), (dii), and (diii), respectively. The dark concentrations in Fig.~\ref{fig:05}~(di) follow the orbit of family $\mathcal{A}$ ($\widetilde{\mathcal{A}}$) at energy $\epsilon=-0.5$ shown with a blue solid (dashed) line. The concentrations in  Fig.~\ref{fig:05}~(dii) follow the orbit from $\mathcal{B}$ ($\widetilde{\mathcal{B}}$) at the same energy shown with a solid (dashed) red line. Interestingly, some dark concentrations are visible also in Fig. \ref{fig:05} (diii), which suggest that state $\ket{\bm x_\text{iii}}$ is scarred, but by POs from families other than those identified in this work.

The fact that the survival probability of state $\ket{\bm x_\text{iii}}$ is so well described by random matrix theory [green line in Fig.~\ref{fig:05}~(aiii) and Fig.~\ref{fig:05}~(biii)] and no identifiable structure is visible in the autocorrelation of its LDoS [Fig.~\ref{fig:05}~(ciii) and Eq. \eqref{eqn:inverseFourierofSP})] suggests that even though this state has a minor degree of scarring that is visible in phase space [Fig. \ref{fig:05} (diii)], it is not identifiable by just looking in the Hilbert space.  To better understand this feature, we use the ratio between the asymptotic value of the survival probability $S_P^\infty$ of $\ket{\bm x_\text{iii}}$ and the asymtpotic value of the survival probability of an ensemble of random states with the same enveloping LDoS $S_P^{(r),\infty}$ \cite{Villasenor2020},
\begin{equation}
R=\frac{S_P^{(r),\infty}}{S_P^\infty}.
\end{equation}
The ratio $R$ measures the similarity between the LDoS of the coherent state and that of a random state, (see Ref.~\cite{Villasenor2020} for an extensive discussion of $R$). A ratio $R=1$ indicates that these distributions are very similar. Lower values imply that there are periodic structures in the LDoS of the chosen coherent states, which are, evidently, absent in the LDoS of a random state~\cite{Lerma2019}.
For states $\ket{\bm x_\text{i}}$  and $\ket{\bm x_\text{ii}}$,  $R_\text{i}=0.41$ and $R_\text{ii}=0.81$, respectively, while $R_\text{iii}=1.01$ for state $\ket{\bm x_\text{iii}}$. These numbers show that the scarring can be traced back to a comb-like structure in the LDoS of states $\ket{\bm x_\text{i}}$  and $\ket{\bm x_\text{iii}}$, but not of state $\ket{\bm x_\text{iii}}$. This is consistent with Ref. \cite{PilatowskyARXIV}, where we found concentrations resembling dynamical scars in the phase-space projections of even the most random-like coherent states at energy $\epsilon=-0.5$, which display no revivals, no comb-like structures in their LDoS, and a path to equilibrium well-described by random matrix theory. 

The dynamical behavior of the survival probability of the coherent states can be described by two competing effects. Periodic structures in the LDoS give rise to revivals, while random-like spreading within the Gaussian envelope of the LDoS prevents revivals. In between the two cases, one may find coherent states whose LDoS display a slightly larger participation of some eigenstates separated by a nearly constant energy difference, but this periodic structure does not stand out significantly over the Gaussian envelope, so revivals are not observed. Yet, when one performs infinite-time averages, the scars corresponding to the high participating eigenstates become visible as in Fig. \ref{fig:05} (diii). A complete description of this kind of coherent states requires the challenging task of identifying the family, or set of families, of POs that scar those high participating eigenstates, so that one can compare their Lyapunov times and periods. This is an interesting idea for a future work.


\section{CONCLUSIONS}
\label{sec:conclusions}

We identified the two fundamental families of classical periodic orbits (POs) that emanate from the ground state of the Dicke model in the superradiant phase and extensively explored their effects in the quantum domain. By introducing a measure of scarring based on the temporal average of the Husimi function of an eigenstate along a PO, we were able to identify which eigenstates are scarred by which family of POs, and by projecting the Husimi functions over the energy shell corresponding to the energy of the eigenstate, we found an effective way to visualize the concentrations around the POs of those families.  

 We also showed that knowledge of the periods and Lyapunov exponents of the POs in the two identified families allows us to characterize the distribution of scarred eigenstates. The energies of the eigenstates scarred by the two families cluster around values obtained by means of the Bohr-Sommerfeld quantization rule of the periods, and the width of these clusters is directly related to the Lyapunov exponents of the POs generating the scars. 

The dynamical consequences of the presence of eigenstates scarred by the two identified families  were studied by considering the survival probability of three representative initial coherent states, two localized in the vicinity of the POs of the two families and one away from them. For the two states close to the POs, our detailed knowledge of the families allowed us to understand the revivals, their periods, and the saturation values of the survival probability. By performing the infinite-time averages of the density matrices of these initial coherent states, dynamical scars were observed. The third initial state shows a survival probability and a local density of states akin to those of random initial states, even though the dynamics has contributions from eigenstates scarred by families not identified here. The fact that the potential families associated with these scars are unknown to us prevents us from making conclusive statements about this state. 

An important extension of the present study would be the systematic identification of more families of POs and the analysis of how they influence the spectrum and dynamics of the model.


\section*{ACKNOWLEDGMENTS}

We thank D. Wisniacki for his valuable comments and acknowledge the support of the Computation Center - ICN, in particular of Enrique Palacios, Luciano D\'iaz, and Eduardo Murrieta. SP-C, DV and JGH acknowledge financial support from the DGAPA- UNAM project IN104020, and SL-H from the Mexican CONACyT project CB2015-01/255702. LFS was supported by the NSF grant No. DMR-1936006.


\appendix
\section{Algorithm to find families of periodic orbits emanating from stationary stable points}
\label{app:findingPOs}

\newcommand{\flow}[1]{
  {\bm \varphi}^{#1}}
Given a stable stationary point $\bm x_\text{GS} \in \mathcal{M}$ with energy $\epsilon_\text{GS}$ and normal period $T_{\epsilon_\text{GS}}$, we iteratively find a continuous family of POs $\mathcal{O}_\epsilon$ starting from $\mathcal{O}_\epsilon=\{\bm x_\text{GS}\}$. The existence of these families is guaranteed by theorem 2.1 of Ref.~\cite{Weinstein1973}.

First, we detail a variant of an algorithm known as the monodromy method~\cite{Baranger1988,DeAguiar1988}. This is a Newton-Raphson-type algorithm that converges towards a PO given an initial guess for an initial condition and period. This type of algorithms has been extensively studied in several systems. See, for example, Refs.~\cite{Simonovi1999,Deaguiar1992}. Then, we detail an algorithm to iteratively construct the guesses required to find the POs.

\subsection{The monodromy method: converging to a periodic orbit given a good initial guess}

Assume we have guesses $\check{\bm{x}}$ and $\check {T}$ for an initial condition and period of a PO, respectively. We want $\bm {x}=\check{\bm {x} }+ \Delta {\bm x}$ and $T=\check{T} + \Delta T$, the initial condition and period of an orbit in the same energy shell as $\check{\bm{x}}$. Denote by $\bm{\Phi}_{\bm x}(t)$ the fundamental matrix associated to the Hamiltonian system $h_\text{cl}$ and $\flow{t}(\bm x)$ the Hamiltonian flow satisfying $\flow{t}(\bm x)=\bm x(t)$ (See \cite{Gaspard1998} 1.1.3). If $\norm{\Delta {\bm x}}$ is small, one may approximate to first order,
 \begin{equation*}
 \flow{\check{T}}(\check{\bm x} + \Delta {\bm x})\approx\flow{\check{T}}(\check{\bm x}) + \bm{\Phi}_{\check{\bm x}}(\check{T})\Delta {\bm x}. \\ 
 \end{equation*}
  
Similarly, if $\abs{\Delta T}$ is small, Taylor expanding the flow up to first order,  
 \begin{align*}
 \flow{\Delta T}(\bm x)\approx \bm x + \Delta T\, \bm \Sigma \nabla h_\text{cl} (\bm x), & \quad&\bm\Sigma=\begin{pmatrix}
0&1&0&0\\
-1&0&0&0\\
0&0&0&1\\
0&0&-1&0 \\
\end{pmatrix}, 
 \end{align*}
 where $\nabla h_\text{cl} (\bm x)$ is the gradient of the Hamiltonian. Then, we have
\begin{align}
\flow{T}(\bm {x})&=\flow{\Delta T+\check{T}}(\bm {x})   
	=\flow{\Delta T}\left(\flow{\check{T}}(\check{\bm x} + \Delta {\bm x})\right) \\ \nonumber
	&\approx\flow{\check{T}}(\check{\bm x} + \Delta {\bm x}) + \Delta T\, \bm \Sigma \nabla h_\text{cl} \left(\flow{\check{T}}(\check{\bm x} + \Delta {\bm x}) \right) \\  \nonumber	
	&\approx\flow{\check{T}}(\check{\bm x}) + \bm{\Phi}_{\check{\bm x}}(\check{T})\Delta {\bm x} +\Delta T\, \bm \Sigma \nabla h_\text{cl} \left(\flow{\check{T}}(\check{\bm x}) + \bm{\Phi}_{\check{\bm x}}(\check{T})\Delta {\bm x} \right) \\  \nonumber
	&\approx \flow{\check{T}}(\check{\bm x}) + \bm{\Phi}_{\check{\bm x}}(\check{T})\Delta {\bm x} + \Delta T\, \bm \Sigma \nabla h_\text{cl} \left(\check{\bm x}' \right) \nonumber
\end{align}
where $\check{\bm x}'=\flow{\check{T}}(\check{\bm x})$. Thus, we can approximate the periodicity constriction $\bm {x}=\flow{T}(\bm {x})$ to first order by
\begin{equation}
\label{eqn:AppPOAlgConst1}
\check{\bm x} + \Delta {\bm x}= \flow{\check{T}}(\check{\bm x}) + \bm{\Phi}_{\check{\bm x}}(\check{T})\Delta {\bm x} + \Delta T\, \bm \Sigma \nabla h_\text{cl} \left(\check{\bm x}' \right).
\end{equation}
 Also, we can approximate the energy constriction $h_\text{cl}(\bm x)=h_\text{cl}(\check{\bm x})$ to first order to get
 \begin{equation}
 \label{eqn:AppPOAlgConst2}
\nabla h_\text{cl}\left(\check{\bm x}' \right) \cdot \Delta {\bm x} =0,
  \end{equation}
and, finally, the constriction to stay in the same Poincar\'e section of constant $P$ as
 \begin{equation}
 \label{eqn:AppPOAlgConst3}
\bm \xi \cdot \Delta {\bm x} =0,
  \end{equation}
where $\bm \xi=(q=0,p=0;Q=0,P=1)^\top$. This last constriction eliminates movement along the flow and increases the stability of the algorithm.

The linear constrictions~\eqref{eqn:AppPOAlgConst1}, \eqref{eqn:AppPOAlgConst2} and \eqref{eqn:AppPOAlgConst3}, may be written in matrix form as
 \begin{equation*}
 \renewcommand\arraystretch{1.8}
 \begin{pmatrix}
 \left(\bm 1-\bm{\Phi}_{\check{\bm x}}(\check{T})\right) & - \bm \Sigma \nabla h_\text{cl}\left(\check{\bm x}' \right) \\
 \nabla h_\text{cl}\left(\check{\bm x}' \right)^\top & 0 \\
  {\bm \xi}^\top & 0

 \end{pmatrix}
 \begin{pmatrix}
\Delta \bm x \\ \\ \Delta T 
 \end{pmatrix}
 =
 \begin{pmatrix}
\check{\bm x}' - \check{\bm x} \\0 \\ 0 
 \end{pmatrix}.
 \end{equation*}
This overdetermined system of linear equations may be approximately solved by least squares using Moore-Penrose pseudoinversion. The solution for $\Delta \bm x$ and $\Delta T$ is not exact, but the process may be iterated with new guesses $\check{\bm x}+\Delta \bm x$ and $\check{T}+\Delta T$ that will converge to $\bm {x}$ and $T$ if the initial guesses were good enough.

\subsection{Finding families of periodic orbits}

We now define an iterative process to find the families of POs. 
For the first step, we start with a stable stationary point $\bm x_\text{GS} \in \mathcal{M}$ with energy $\epsilon_\text{GS}=h_\text{cl}(\bm x_\text{GS})$ and normal period $T_{\epsilon_\text{GS}}$.

Given $\mathcal{O}_{\epsilon}$, we now explain how to find $\mathcal{O}_{\epsilon'}$ with $\epsilon' = \delta \epsilon + \epsilon$ close to $\epsilon$. Pick $\bm x \in \mathcal{O}$ and define a perturbation $\delta \bm x= a \nabla h_{\text{cl}}(\bm x)$, where $a$ is a scalar such that ${h_\text{cl}(\bm x +\delta \bm x)= \epsilon +\delta \epsilon}$. For the first step, $\nabla h_{\text{cl}}(\bm x_\text{GS})=0$, in which case one may select any direction (we use the $q$ direction), and the stability of the initial stationary point will guarantee that the algorithm converges.
Set
\begin{align}
&&\check{\bm x}'=\bm x +\delta \bm x &&\text{and}&&
\check{T}' = T+ \delta T, &&
\end{align}
where 
\begin{equation*}
\delta T_{k}=\begin{cases}
0 & \text{ if we are in the first step,} \\
(\epsilon'-\epsilon){T - T_\text{prev}\over \epsilon - \epsilon_\text{prev}} & \text{else,} \\ \end{cases}
\end{equation*}
where $\epsilon_\text{prev}$ and $T_\text{prev}$ are the energy and period of the closest previously calculated orbit. This way, $\check{T}'$ is linear extrapolation based on the behavior of the previous orbits.

Using the monodromy method detailed in the previous subsection, we may correct the guesses $\check{\bm x}'$ and $\check{T}'$ to obtain actual solutions $\bm {x}'$ and $T'$ so that the desired PO is
\begin{equation}
\mathcal{O}_{\epsilon'} = \left \{\flow{t}(\bm {x}') \, \vert \, t\in [0, T'] \right \}.
\end{equation}
Although energy is constrained in the monodromy method we used, this is only to first order, so $\epsilon'=h_\text{cl}(\bm x')$ may not be exactly equal to $\delta \epsilon + \epsilon$. This is easily fixed by performing additional iterations with smaller $\abs{\delta\epsilon}$ which converge to the desired energy.

\section{Computation of the Lyapunov exponent of a periodic orbit}
\label{app:Lyapunovs}
For any point of the phase space $\bm x$, the associated maximal Lyapunov exponent $\lambda$ may be calculated with the spectral norm of the fundamental matrix, $\norm{\bm{\Phi}_{\bm x}(t)}$, which is the square root of the maximal eigenvalue of the symmetric matrix $\bm{\Phi}_{\bm x}(t)^\dagger\bm{\Phi}_{\bm x}(t)$. The formula reads \cite{Gaspard1998,Chavez2019,Pilatowsky2020} 
\begin{equation}
\label{eqn:lyapunovdefinition}
\lambda= \lim_{t\to \infty} \frac{1}{t} \log \norm{\bm{\Phi}_{\bm x}(t)}.
\end{equation}
In the case that $\bm x$ corresponds to a periodic condition of period $T$, Eq.~\eqref{eqn:lyapunovdefinition} is greatly simplified. The so-called monodromy matrix associated to $\bm x$ is $M=\bm{\Phi}_{\bm x}(T)$. Then $\bm{\Phi}_{\bm x}(n T)=M^n$ for all positive integers $n$ \cite{Gaspard1998}. Thus,
\begin{equation}
\label{eqn:lyaexpansion}
\lambda= \lim_{n\to \infty} \frac{1}{nT} \log \norm{\bm{\Phi}_{\bm x}(nT)}=\lim_{n\to \infty} \frac{1}{nT} \log \norm{M^n}=\frac{1}{T}\lim_{n\to \infty} \frac{1}{n} \log \norm{e^{nA}},
\end{equation}
where $A=\log M$. In Ref. \cite{Pilatowsky2020}, it was shown that the rightmost limit of  Eq.~\eqref{eqn:lyaexpansion} is equal to $\max_i \Re(a_i)$. That is the maximal real part of the eigenvalues $a_i$ of $A$, which are given by $a_i=\log m_i$, where $m_i$ are the eigenvalues of $M$.  Thus, 
\begin{equation}
\lambda=\frac{1}{T}\max_i \Re(\log m_i)=\frac{1}{T}\max_i \log \abs{m_i}.
\end{equation}

In short, finding the maximal Lyapunov exponent of a periodic orbit of period $T$ reduces to computing the monodromy matrix $M$, which is done by simple numeric integration, and then taking the greatest of the norms of its eigenvalues divided by $T$.

\section{The Husimi distribution along periodic orbits}
\label{app:ScarringAsOverlap}

The unnormalized Husimi function for a coherent state $\ket{\bm y}$, as given by Eq.~\eqref{eq:Hu_pure_state}, is very well fitted by 
\begin{equation}
\mathcal{Q}_{\bm y}(\bm x)=\exp(-\frac{j}{2} \,d_\mathcal{M}(\bm x,\bm y)^2),
\label{eqn:husimi_coherent}
\end{equation}
for values of $j$ larger than $\sim 10$. The distance $d_\mathcal{M}$ reads
\begin{equation*}
d_\mathcal{M}\big((q,p,\theta,\phi),(q',p',\theta',\phi')\big)^2=(q-q')^2 + (p - p')^2 + \Theta^2,
\end{equation*}
where 
$\cos\Theta=\cos\theta\cos\theta' +  \cos(\phi - \phi')\sin\theta\sin\theta'$, and $(\theta, \phi)$ are the spherical coordinates for the Bloch sphere, $\tan\phi=-P/Q$ and $\cos\theta=1 - (Q^2 + P^2)/2$.
Inserting Eq.~\eqref{eqn:husimi_coherent} into Eq.~\eqref{eqn:rho_O}, we get
\begin{equation}
\label{eq:Q_rho_O}
\mathcal{Q}_{\rho_\mathcal{O}} (\bm x)  = \bra{\bm x} \rho_{\mathcal{O}} \ket{\bm x} =\frac{1}{ T}\int_0^T \dif  t \, \exp(-\frac{j}{2} \, d_\mathcal{M}\big({\bm y}(t),\bm x\big)^2), 
\end{equation}
for any initial $\bm y \in \mathcal{O}$.

\begin{figure}[ht]
\centering
\includegraphics[width=0.5\columnwidth]{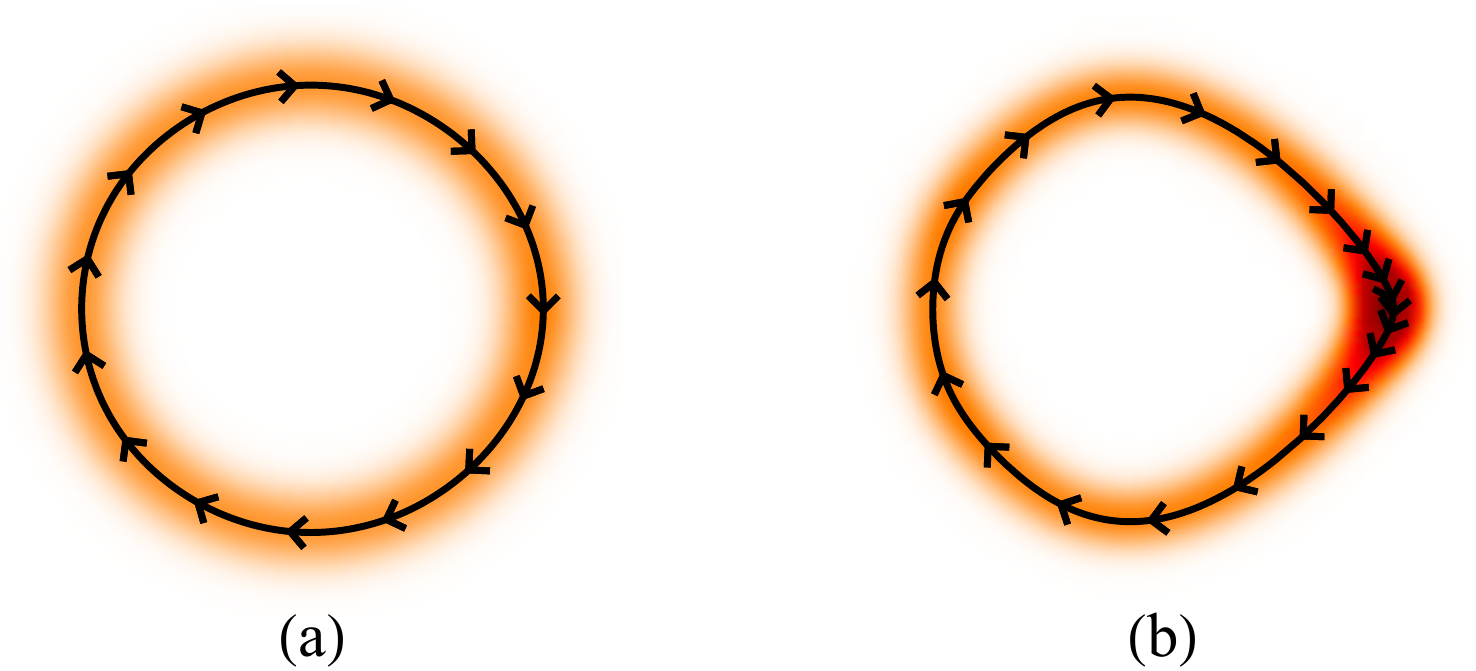} 
\caption{Periodic orbit (black solid line) $\mathcal{O}$ and the Husimi function $\mathcal{Q}_\mathcal{O}$ [Eq.~\eqref{eq:Q_rho_O}] of the corresponding distribution $\hat{\rho}_\mathcal{O}$ (color). Darker color indicates a higher value of the Husimi function. The arrows are placed at intervals of constant time, so closer arrows indicate slower dynamics.} 
\label{fig:06}
\end{figure}

As an illustration, we plot in  Fig.~\ref{fig:06} the Husimi function of $\hat{\rho}_\mathcal{O}$ for two generic POs.

\section{Computation of projected Husimi distribution}
\label{app:ProjHusimi}

Because of the properties of the $\delta$ function, Eq.~\eqref{eqn:ProjHusimiDef} equals
\begin{equation}
\label{eqn:NumHu}
\widetilde{\mathcal{Q}}_{\epsilon,\hat{\rho}}(Q,P) = \int_{p_{-}}^{p_{+}}  \dif p \frac{\sum_{q_{\pm}}\mathcal{Q}_{\hat{\rho}}(q_{\pm},p;Q,P)}{\sqrt{\Delta(\epsilon_,p,Q,P)}},
\end{equation}
where $q_{\pm}$ are the two solutions in $q$ of the second-degree equation $h_\text{cl}(q,p;Q,P)=\epsilon$,  $p_{\pm}$ are the two solutions in $p$ of the second-degree equation $\Delta(\epsilon,p,Q,P)=0$, and
\begin{align}
\Delta(\epsilon,p,Q,P)  = \abs{\frac{\partial h_\text{cl}}{\partial q} (q_\pm,p;Q,P)}^{2} 
 = 2 \omega \omega_{0} \left(\frac{\epsilon}{\omega_{0}} + 1 - \frac{Q^{2} + P^{2}}{2}\right) + 
 4 \gamma^{2} Q^{2} \left(1 - \frac{Q^{2} + P^{2}}{4}\right) - \omega^{2} p^{2}.
\end{align}
 We compute the integral~\eqref{eqn:NumHu}  with a Chebyshev-Gauss quadrature method. \cite{PilatowskyARXIV}



\begin{thebibliography}{71}%
\makeatletter
\providecommand \@ifxundefined [1]{%
 \@ifx{#1\undefined}
}%
\providecommand \@ifnum [1]{%
 \ifnum #1\expandafter \@firstoftwo
 \else \expandafter \@secondoftwo
 \fi
}%
\providecommand \@ifx [1]{%
 \ifx #1\expandafter \@firstoftwo
 \else \expandafter \@secondoftwo
 \fi
}%
\providecommand \natexlab [1]{#1}%
\providecommand \enquote  [1]{``#1''}%
\providecommand \bibnamefont  [1]{#1}%
\providecommand \bibfnamefont [1]{#1}%
\providecommand \citenamefont [1]{#1}%
\providecommand \href@noop [0]{\@secondoftwo}%
\providecommand \href [0]{\begingroup \@sanitize@url \@href}%
\providecommand \@href[1]{\@@startlink{#1}\@@href}%
\providecommand \@@href[1]{\endgroup#1\@@endlink}%
\providecommand \@sanitize@url [0]{\catcode `\\12\catcode `\$12\catcode
  `\&12\catcode `\#12\catcode `\^12\catcode `\_12\catcode `\%12\relax}%
\providecommand \@@startlink[1]{}%
\providecommand \@@endlink[0]{}%
\providecommand \url  [0]{\begingroup\@sanitize@url \@url }%
\providecommand \@url [1]{\endgroup\@href {#1}{\urlprefix }}%
\providecommand \urlprefix  [0]{URL }%
\providecommand \Eprint [0]{\href }%
\providecommand \doibase [0]{http://dx.doi.org/}%
\providecommand \selectlanguage [0]{\@gobble}%
\providecommand \bibinfo  [0]{\@secondoftwo}%
\providecommand \bibfield  [0]{\@secondoftwo}%
\providecommand \translation [1]{[#1]}%
\providecommand \BibitemOpen [0]{}%
\providecommand \bibitemStop [0]{}%
\providecommand \bibitemNoStop [0]{.\EOS\space}%
\providecommand \EOS [0]{\spacefactor3000\relax}%
\providecommand \BibitemShut  [1]{\csname bibitem#1\endcsname}%
\let\auto@bib@innerbib\@empty
\bibitem [{\citenamefont {Heller}(1984)}]{Heller1984}%
  \BibitemOpen
  \bibfield  {author} {\bibinfo {author} {\bibfnamefont {Eric~J.}\ \bibnamefont
  {Heller}},\ }\bibfield  {title} {\enquote {\bibinfo {title} {Bound-state
  eigenfunctions of classically chaotic {H}amiltonian systems: Scars of
  periodic orbits},}\ }\href {\doibase 10.1103/PhysRevLett.53.1515} {\bibfield
  {journal} {\bibinfo  {journal} {Phys. Rev. Lett.}\ }\textbf {\bibinfo
  {volume} {53}},\ \bibinfo {pages} {1515--1518} (\bibinfo {year}
  {1984})}\BibitemShut {NoStop}%
\bibitem [{\citenamefont {Berry}(1989)}]{Berry1989}%
  \BibitemOpen
  \bibfield  {author} {\bibinfo {author} {\bibfnamefont {Michael~Victor}\
  \bibnamefont {Berry}},\ }\bibfield  {title} {\enquote {\bibinfo {title}
  {Quantum scars of classical closed orbits in phase space},}\ }\href {\doibase
  10.1098/rspa.1989.0052} {\bibfield  {journal} {\bibinfo  {journal}
  {Proceedings of the Royal Society of London. A. Mathematical and Physical
  Sciences}\ }\textbf {\bibinfo {volume} {423}},\ \bibinfo {pages} {219--231}
  (\bibinfo {year} {1989})}\BibitemShut {NoStop}%
\bibitem [{\citenamefont {Heller}(1991)}]{Heller1991}%
  \BibitemOpen
  \bibfield  {author} {\bibinfo {author} {\bibfnamefont {E.~J.}\ \bibnamefont
  {Heller}},\ }\bibfield  {title} {\enquote {\bibinfo {title} {Wavepacket
  dynamics and quantum chaology},}\ }in\ \href@noop {} {\emph {\bibinfo
  {booktitle} {Les Houches Summer School 1991 on Chaos and Quantum Physics}}},\
  \bibinfo {editor} {edited by\ \bibinfo {editor} {\bibfnamefont {M.-J.}\
  \bibnamefont {Giannoni}}, \bibinfo {editor} {\bibfnamefont {A.}~\bibnamefont
  {Voros}}, \ and\ \bibinfo {editor} {\bibfnamefont {J.~Zinn}\ \bibnamefont
  {Justin}}}\ (\bibinfo  {publisher} {Springer},\ \bibinfo {year}
  {1991})\BibitemShut {NoStop}%
\bibitem [{\citenamefont {McDonald}\ and\ \citenamefont
  {Kaufman}(1979)}]{McDonald1979}%
  \BibitemOpen
  \bibfield  {author} {\bibinfo {author} {\bibfnamefont {Steven~W.}\
  \bibnamefont {McDonald}}\ and\ \bibinfo {author} {\bibfnamefont {Allan~N.}\
  \bibnamefont {Kaufman}},\ }\bibfield  {title} {\enquote {\bibinfo {title}
  {Spectrum and eigenfunctions for a {H}amiltonian with stochastic
  trajectories},}\ }\href {\doibase 10.1103/PhysRevLett.42.1189} {\bibfield
  {journal} {\bibinfo  {journal} {Phys. Rev. Lett.}\ }\textbf {\bibinfo
  {volume} {42}},\ \bibinfo {pages} {1189--1191} (\bibinfo {year}
  {1979})}\BibitemShut {NoStop}%
\bibitem [{\citenamefont {St\"ockmann}(2006)}]{StockmannBook}%
  \BibitemOpen
  \bibfield  {author} {\bibinfo {author} {\bibfnamefont {H-J}\ \bibnamefont
  {St\"ockmann}},\ }\href@noop {} {\emph {\bibinfo {title} {Quantum Chaos: An
  Introduction}}}\ (\bibinfo  {publisher} {Cambridge University Press},\
  \bibinfo {address} {Cambridge},\ \bibinfo {year} {2006})\BibitemShut
  {NoStop}%
\bibitem [{\citenamefont {Wintgen}\ and\ \citenamefont
  {H\"{o}nig}(1989)}]{Wintgen1989}%
  \BibitemOpen
  \bibfield  {author} {\bibinfo {author} {\bibfnamefont {D.}~\bibnamefont
  {Wintgen}}\ and\ \bibinfo {author} {\bibfnamefont {A.}~\bibnamefont
  {H\"{o}nig}},\ }\bibfield  {title} {\enquote {\bibinfo {title} {Irregular
  wave functions of a hydrogen atom in a uniform magnetic field},}\ }\href
  {\doibase 10.1103/physrevlett.63.1467} {\bibfield  {journal} {\bibinfo
  {journal} {Phys. Rev. Lett.}\ }\textbf {\bibinfo {volume} {63}},\ \bibinfo
  {pages} {1467--1470} (\bibinfo {year} {1989})}\BibitemShut {NoStop}%
\bibitem [{\citenamefont {D'Ariano}\ \emph {et~al.}(1992)\citenamefont
  {D'Ariano}, \citenamefont {Evangelista},\ and\ \citenamefont
  {Saraceno}}]{Ariano1992}%
  \BibitemOpen
  \bibfield  {author} {\bibinfo {author} {\bibfnamefont {G.~M.}\ \bibnamefont
  {D'Ariano}}, \bibinfo {author} {\bibfnamefont {L.~R.}\ \bibnamefont
  {Evangelista}}, \ and\ \bibinfo {author} {\bibfnamefont {M.}~\bibnamefont
  {Saraceno}},\ }\bibfield  {title} {\enquote {\bibinfo {title} {Classical and
  quantum structures in the kicked-top model},}\ }\href {\doibase
  10.1103/PhysRevA.45.3646} {\bibfield  {journal} {\bibinfo  {journal} {Phys.
  Rev. A}\ }\textbf {\bibinfo {volume} {45}},\ \bibinfo {pages} {3646--3658}
  (\bibinfo {year} {1992})}\BibitemShut {NoStop}%
\bibitem [{\citenamefont {Heller}(1987)}]{Heller1987}%
  \BibitemOpen
  \bibfield  {author} {\bibinfo {author} {\bibfnamefont {Eric~J.}\ \bibnamefont
  {Heller}},\ }\bibfield  {title} {\enquote {\bibinfo {title} {Quantum
  localization and the rate of exploration of phase space},}\ }\href {\doibase
  10.1103/PhysRevA.35.1360} {\bibfield  {journal} {\bibinfo  {journal} {Phys.
  Rev. A}\ }\textbf {\bibinfo {volume} {35}},\ \bibinfo {pages} {1360--1370}
  (\bibinfo {year} {1987})}\BibitemShut {NoStop}%
\bibitem [{\citenamefont {Bogomolny}(1988)}]{Bogomolny1988}%
  \BibitemOpen
  \bibfield  {author} {\bibinfo {author} {\bibfnamefont {E.B.}\ \bibnamefont
  {Bogomolny}},\ }\bibfield  {title} {\enquote {\bibinfo {title} {Smoothed wave
  functions of chaotic quantum systems},}\ }\href {\doibase
  https://doi.org/10.1016/0167-2789(88)90075-9} {\bibfield  {journal} {\bibinfo
   {journal} {Phys. D}\ }\textbf {\bibinfo {volume} {31}},\ \bibinfo {pages}
  {169 -- 189} (\bibinfo {year} {1988})}\BibitemShut {NoStop}%
\bibitem [{\citenamefont {Agam}\ and\ \citenamefont
  {Fishman}(1993)}]{Agam1993}%
  \BibitemOpen
  \bibfield  {author} {\bibinfo {author} {\bibfnamefont {O}~\bibnamefont
  {Agam}}\ and\ \bibinfo {author} {\bibfnamefont {S}~\bibnamefont {Fishman}},\
  }\bibfield  {title} {\enquote {\bibinfo {title} {Quantum eigenfunctions in
  terms of periodic orbits of chaotic systems},}\ }\href {\doibase
  10.1088/0305-4470/26/9/010} {\bibfield  {journal} {\bibinfo  {journal} {J.
  Phys. A}\ }\textbf {\bibinfo {volume} {26}},\ \bibinfo {pages} {2113--2137}
  (\bibinfo {year} {1993})}\BibitemShut {NoStop}%
\bibitem [{\citenamefont {Bohigas}\ \emph {et~al.}(1993)\citenamefont
  {Bohigas}, \citenamefont {Tomsovic},\ and\ \citenamefont
  {Ullmo}}]{Bohigas1993}%
  \BibitemOpen
  \bibfield  {author} {\bibinfo {author} {\bibfnamefont {O.}~\bibnamefont
  {Bohigas}}, \bibinfo {author} {\bibfnamefont {S.}~\bibnamefont {Tomsovic}}, \
  and\ \bibinfo {author} {\bibfnamefont {D.}~\bibnamefont {Ullmo}},\ }\bibfield
   {title} {\enquote {\bibinfo {title} {Manifestations of classical phase space
  structures in quantum mechanics},}\ }\href {\doibase
  10.1016/0370-1573(93)90109-q} {\bibfield  {journal} {\bibinfo  {journal}
  {Phys. Rep.}\ }\textbf {\bibinfo {volume} {223}},\ \bibinfo {pages} {43--133}
  (\bibinfo {year} {1993})}\BibitemShut {NoStop}%
\bibitem [{\citenamefont {Muller}\ and\ \citenamefont
  {Wintgen}(1994)}]{Muller1994}%
  \BibitemOpen
  \bibfield  {author} {\bibinfo {author} {\bibfnamefont {K}~\bibnamefont
  {Muller}}\ and\ \bibinfo {author} {\bibfnamefont {D}~\bibnamefont
  {Wintgen}},\ }\bibfield  {title} {\enquote {\bibinfo {title} {Scars in
  wavefunctions of the diamagnetic kepler problem},}\ }\href {\doibase
  10.1088/0953-4075/27/13/003} {\bibfield  {journal} {\bibinfo  {journal} {J.
  Phys. B}\ }\textbf {\bibinfo {volume} {27}},\ \bibinfo {pages} {2693--2718}
  (\bibinfo {year} {1994})}\BibitemShut {NoStop}%
\bibitem [{\citenamefont {Kaplan}\ and\ \citenamefont
  {Heller}(1998)}]{Kaplan1998}%
  \BibitemOpen
  \bibfield  {author} {\bibinfo {author} {\bibfnamefont {L.}~\bibnamefont
  {Kaplan}}\ and\ \bibinfo {author} {\bibfnamefont {E.J.}\ \bibnamefont
  {Heller}},\ }\bibfield  {title} {\enquote {\bibinfo {title} {Linear and
  nonlinear theory of eigenfunction scars},}\ }\href {\doibase
  https://doi.org/10.1006/aphy.1997.5773} {\bibfield  {journal} {\bibinfo
  {journal} {Ann. of Phys.}\ }\textbf {\bibinfo {volume} {264}},\ \bibinfo
  {pages} {171 -- 206} (\bibinfo {year} {1998})}\BibitemShut {NoStop}%
\bibitem [{\citenamefont {Kaplan}\ and\ \citenamefont
  {Heller}(1999)}]{Kaplan1999}%
  \BibitemOpen
  \bibfield  {author} {\bibinfo {author} {\bibfnamefont {L.}~\bibnamefont
  {Kaplan}}\ and\ \bibinfo {author} {\bibfnamefont {E.~J.}\ \bibnamefont
  {Heller}},\ }\bibfield  {title} {\enquote {\bibinfo {title} {Measuring scars
  of periodic orbits},}\ }\href {\doibase 10.1103/PhysRevE.59.6609} {\bibfield
  {journal} {\bibinfo  {journal} {Phys. Rev. E}\ }\textbf {\bibinfo {volume}
  {59}},\ \bibinfo {pages} {6609--6628} (\bibinfo {year} {1999})}\BibitemShut
  {NoStop}%
\bibitem [{\citenamefont {Wisniacki}\ \emph {et~al.}(2006)\citenamefont
  {Wisniacki}, \citenamefont {Vergini}, \citenamefont {Benito},\ and\
  \citenamefont {Borondo}}]{Wisniacki2006}%
  \BibitemOpen
  \bibfield  {author} {\bibinfo {author} {\bibfnamefont {D.~A.}\ \bibnamefont
  {Wisniacki}}, \bibinfo {author} {\bibfnamefont {E.}~\bibnamefont {Vergini}},
  \bibinfo {author} {\bibfnamefont {R.~M.}\ \bibnamefont {Benito}}, \ and\
  \bibinfo {author} {\bibfnamefont {F.}~\bibnamefont {Borondo}},\ }\bibfield
  {title} {\enquote {\bibinfo {title} {Scarring by homoclinic and heteroclinic
  orbits},}\ }\href {\doibase 10.1103/PhysRevLett.97.094101} {\bibfield
  {journal} {\bibinfo  {journal} {Phys. Rev. Lett.}\ }\textbf {\bibinfo
  {volume} {97}},\ \bibinfo {pages} {094101} (\bibinfo {year}
  {2006})}\BibitemShut {NoStop}%
\bibitem [{\citenamefont {Porter}\ \emph {et~al.}(2017)\citenamefont {Porter},
  \citenamefont {Barr}, \citenamefont {Barr},\ and\ \citenamefont
  {Reichl}}]{Porter2017}%
  \BibitemOpen
  \bibfield  {author} {\bibinfo {author} {\bibfnamefont {Max~D.}\ \bibnamefont
  {Porter}}, \bibinfo {author} {\bibfnamefont {Aaron}\ \bibnamefont {Barr}},
  \bibinfo {author} {\bibfnamefont {Ariel}\ \bibnamefont {Barr}}, \ and\
  \bibinfo {author} {\bibfnamefont {L.~E.}\ \bibnamefont {Reichl}},\ }\bibfield
   {title} {\enquote {\bibinfo {title} {{Chaos in the band structure of a soft
  Sinai lattice}},}\ }\href {\doibase 10.1103/PhysRevE.95.052213} {\bibfield
  {journal} {\bibinfo  {journal} {Phys. Rev. E}\ }\textbf {\bibinfo {volume}
  {95}},\ \bibinfo {pages} {052213} (\bibinfo {year} {2017})}\BibitemShut
  {NoStop}%
\bibitem [{\citenamefont {Keski-Rahkonen}\ \emph
  {et~al.}(2019{\natexlab{a}})\citenamefont {Keski-Rahkonen}, \citenamefont
  {Luukko}, \citenamefont {{\AA}berg},\ and\ \citenamefont
  {Räsänen}}]{Keski2019}%
  \BibitemOpen
  \bibfield  {author} {\bibinfo {author} {\bibfnamefont {J}~\bibnamefont
  {Keski-Rahkonen}}, \bibinfo {author} {\bibfnamefont {P~J~J}\ \bibnamefont
  {Luukko}}, \bibinfo {author} {\bibfnamefont {S}~\bibnamefont {{\AA}berg}}, \
  and\ \bibinfo {author} {\bibfnamefont {E}~\bibnamefont {Räsänen}},\
  }\bibfield  {title} {\enquote {\bibinfo {title} {Effects of scarring on
  quantum chaos in disordered quantum wells},}\ }\href {\doibase
  10.1088/1361-648x/aaf9fb} {\bibfield  {journal} {\bibinfo  {journal} {J.
  Phys. C}\ }\textbf {\bibinfo {volume} {31}},\ \bibinfo {pages} {105301}
  (\bibinfo {year} {2019}{\natexlab{a}})}\BibitemShut {NoStop}%
\bibitem [{\citenamefont {Keski-Rahkonen}\ \emph
  {et~al.}(2019{\natexlab{b}})\citenamefont {Keski-Rahkonen}, \citenamefont
  {Ruhanen}, \citenamefont {Heller},\ and\ \citenamefont
  {R\"as\"anen}}]{KeskiHeller2019}%
  \BibitemOpen
  \bibfield  {author} {\bibinfo {author} {\bibfnamefont {J.}~\bibnamefont
  {Keski-Rahkonen}}, \bibinfo {author} {\bibfnamefont {A.}~\bibnamefont
  {Ruhanen}}, \bibinfo {author} {\bibfnamefont {E.~J.}\ \bibnamefont {Heller}},
  \ and\ \bibinfo {author} {\bibfnamefont {E.}~\bibnamefont {R\"as\"anen}},\
  }\bibfield  {title} {\enquote {\bibinfo {title} {Quantum {L}issajous
  scars},}\ }\href {\doibase 10.1103/PhysRevLett.123.214101} {\bibfield
  {journal} {\bibinfo  {journal} {Phys. Rev. Lett.}\ }\textbf {\bibinfo
  {volume} {123}},\ \bibinfo {pages} {214101} (\bibinfo {year}
  {2019}{\natexlab{b}})}\BibitemShut {NoStop}%
\bibitem [{\citenamefont {Dicke}(1954)}]{Dicke1954}%
  \BibitemOpen
  \bibfield  {author} {\bibinfo {author} {\bibfnamefont {R.~H.}\ \bibnamefont
  {Dicke}},\ }\bibfield  {title} {\enquote {\bibinfo {title} {Coherence in
  spontaneous radiation processes},}\ }\href {\doibase 10.1103/PhysRev.93.99}
  {\bibfield  {journal} {\bibinfo  {journal} {Phys. Rev.}\ }\textbf {\bibinfo
  {volume} {93}},\ \bibinfo {pages} {99} (\bibinfo {year} {1954})}\BibitemShut
  {NoStop}%
\bibitem [{\citenamefont {de~Aguiar}\ \emph {et~al.}(1992)\citenamefont
  {de~Aguiar}, \citenamefont {Furuya}, \citenamefont {Lewenkopf},\ and\
  \citenamefont {Nemes}}]{Deaguiar1992}%
  \BibitemOpen
  \bibfield  {author} {\bibinfo {author} {\bibfnamefont {M.A.M}\ \bibnamefont
  {de~Aguiar}}, \bibinfo {author} {\bibfnamefont {K}~\bibnamefont {Furuya}},
  \bibinfo {author} {\bibfnamefont {C.H}\ \bibnamefont {Lewenkopf}}, \ and\
  \bibinfo {author} {\bibfnamefont {M.C}\ \bibnamefont {Nemes}},\ }\bibfield
  {title} {\enquote {\bibinfo {title} {Chaos in a spin-boson system: Classical
  analysis},}\ }\href {\doibase https://doi.org/10.1016/0003-4916(92)90178-O}
  {\bibfield  {journal} {\bibinfo  {journal} {Ann. of Phys.}\ }\textbf
  {\bibinfo {volume} {216}},\ \bibinfo {pages} {291 -- 312} (\bibinfo {year}
  {1992})}\BibitemShut {NoStop}%
\bibitem [{\citenamefont {Bakemeier}\ \emph {et~al.}(2013)\citenamefont
  {Bakemeier}, \citenamefont {Alvermann},\ and\ \citenamefont
  {Fehske}}]{Bakemeier2013}%
  \BibitemOpen
  \bibfield  {author} {\bibinfo {author} {\bibfnamefont {L.}~\bibnamefont
  {Bakemeier}}, \bibinfo {author} {\bibfnamefont {A.}~\bibnamefont
  {Alvermann}}, \ and\ \bibinfo {author} {\bibfnamefont {H.}~\bibnamefont
  {Fehske}},\ }\bibfield  {title} {\enquote {\bibinfo {title} {Dynamics of the
  {D}icke model close to the classical limit},}\ }\href {\doibase
  10.1103/PhysRevA.88.043835} {\bibfield  {journal} {\bibinfo  {journal} {Phys.
  Rev. A}\ }\textbf {\bibinfo {volume} {88}},\ \bibinfo {pages} {043835}
  (\bibinfo {year} {2013})}\BibitemShut {NoStop}%
\bibitem [{\citenamefont {Turner}\ \emph {et~al.}(2018)\citenamefont {Turner},
  \citenamefont {Michailidis}, \citenamefont {Abanin}, \citenamefont {Serbyn},\
  and\ \citenamefont {Papif{\'a}}}]{Turner2018}%
  \BibitemOpen
  \bibfield  {author} {\bibinfo {author} {\bibfnamefont {C.~J.}\ \bibnamefont
  {Turner}}, \bibinfo {author} {\bibfnamefont {A.~A.}\ \bibnamefont
  {Michailidis}}, \bibinfo {author} {\bibfnamefont {D.~A.}\ \bibnamefont
  {Abanin}}, \bibinfo {author} {\bibfnamefont {M.}~\bibnamefont {Serbyn}}, \
  and\ \bibinfo {author} {\bibfnamefont {Z.}~\bibnamefont {Papif{\'a}}},\
  }\bibfield  {title} {\enquote {\bibinfo {title} {Weak ergodicity breaking
  from quantum many-body scars},}\ }\href {\doibase 10.1038/s41567-018-0137-5}
  {\bibfield  {journal} {\bibinfo  {journal} {Nat. Phys.}\ }\textbf {\bibinfo
  {volume} {14}},\ \bibinfo {pages} {745--749} (\bibinfo {year}
  {2018})}\BibitemShut {NoStop}%
\bibitem [{\citenamefont {Turner}\ \emph {et~al.}(2020)\citenamefont {Turner},
  \citenamefont {Desaules}, \citenamefont {Bull},\ and\ \citenamefont
  {Papić}}]{Turner2020Arxiv}%
  \BibitemOpen
  \bibfield  {author} {\bibinfo {author} {\bibfnamefont {Christopher~J.}\
  \bibnamefont {Turner}}, \bibinfo {author} {\bibfnamefont {Jean-Yves}\
  \bibnamefont {Desaules}}, \bibinfo {author} {\bibfnamefont {Kieran}\
  \bibnamefont {Bull}}, \ and\ \bibinfo {author} {\bibfnamefont {Zlatko}\
  \bibnamefont {Papić}},\ }\href@noop {} {\enquote {\bibinfo {title}
  {Correspondence principle for many-body scars in ultracold {R}ydberg
  atoms},}\ } (\bibinfo {year} {2020}),\ \Eprint
  {http://arxiv.org/abs/2006.13207} {arXiv:2006.13207 [quant-ph]} \BibitemShut
  {NoStop}%
\bibitem [{\citenamefont {Hepp}\ and\ \citenamefont
  {Lieb}(1973{\natexlab{a}})}]{Hepp1973a}%
  \BibitemOpen
  \bibfield  {author} {\bibinfo {author} {\bibfnamefont {Klaus}\ \bibnamefont
  {Hepp}}\ and\ \bibinfo {author} {\bibfnamefont {Elliott~H}\ \bibnamefont
  {Lieb}},\ }\bibfield  {title} {\enquote {\bibinfo {title} {On the
  superradiant phase transition for molecules in a quantized radiation field:
  the {D}icke maser model},}\ }\href {\doibase
  https://doi.org/10.1016/0003-4916(73)90039-0} {\bibfield  {journal} {\bibinfo
   {journal} {Ann. Phys. (N.Y.)}\ }\textbf {\bibinfo {volume} {76}},\ \bibinfo
  {pages} {360 -- 404} (\bibinfo {year} {1973}{\natexlab{a}})}\BibitemShut
  {NoStop}%
\bibitem [{\citenamefont {Hepp}\ and\ \citenamefont
  {Lieb}(1973{\natexlab{b}})}]{Hepp1973b}%
  \BibitemOpen
  \bibfield  {author} {\bibinfo {author} {\bibfnamefont {Klaus}\ \bibnamefont
  {Hepp}}\ and\ \bibinfo {author} {\bibfnamefont {Elliott~H.}\ \bibnamefont
  {Lieb}},\ }\bibfield  {title} {\enquote {\bibinfo {title} {Equilibrium
  statistical mechanics of matter interacting with the quantized radiation
  field},}\ }\href {\doibase 10.1103/PhysRevA.8.2517} {\bibfield  {journal}
  {\bibinfo  {journal} {Phys. Rev. A}\ }\textbf {\bibinfo {volume} {8}},\
  \bibinfo {pages} {2517--2525} (\bibinfo {year}
  {1973}{\natexlab{b}})}\BibitemShut {NoStop}%
\bibitem [{\citenamefont {Wang}\ and\ \citenamefont {Hioe}(1973)}]{Wang1973}%
  \BibitemOpen
  \bibfield  {author} {\bibinfo {author} {\bibfnamefont {Y.~K.}\ \bibnamefont
  {Wang}}\ and\ \bibinfo {author} {\bibfnamefont {F.~T.}\ \bibnamefont
  {Hioe}},\ }\bibfield  {title} {\enquote {\bibinfo {title} {Phase transition
  in the {D}icke model of superradiance},}\ }\href {\doibase
  10.1103/PhysRevA.7.831} {\bibfield  {journal} {\bibinfo  {journal} {Phys.
  Rev. A}\ }\textbf {\bibinfo {volume} {7}},\ \bibinfo {pages} {831--836}
  (\bibinfo {year} {1973})}\BibitemShut {NoStop}%
\bibitem [{\citenamefont {Emary}\ and\ \citenamefont
  {Brandes}(2003{\natexlab{a}})}]{Emary2003}%
  \BibitemOpen
  \bibfield  {author} {\bibinfo {author} {\bibfnamefont {Clive}\ \bibnamefont
  {Emary}}\ and\ \bibinfo {author} {\bibfnamefont {Tobias}\ \bibnamefont
  {Brandes}},\ }\bibfield  {title} {\enquote {\bibinfo {title} {Chaos and the
  quantum phase transition in the {D}icke model},}\ }\href {\doibase
  10.1103/PhysRevE.67.066203} {\bibfield  {journal} {\bibinfo  {journal} {Phys.
  Rev. E}\ }\textbf {\bibinfo {volume} {67}},\ \bibinfo {pages} {066203}
  (\bibinfo {year} {2003}{\natexlab{a}})}\BibitemShut {NoStop}%
\bibitem [{\citenamefont {Garraway}(2011)}]{Garraway2011}%
  \BibitemOpen
  \bibfield  {author} {\bibinfo {author} {\bibfnamefont {Barry~M.}\
  \bibnamefont {Garraway}},\ }\bibfield  {title} {\enquote {\bibinfo {title}
  {The {D}icke model in quantum optics: {D}icke model revisited},}\ }\href
  {\doibase 10.1098/rsta.2010.0333} {\bibfield  {journal} {\bibinfo  {journal}
  {Philos. Trans. Royal Soc. A}\ }\textbf {\bibinfo {volume} {369}},\ \bibinfo
  {pages} {1137} (\bibinfo {year} {2011})}\BibitemShut {NoStop}%
\bibitem [{\citenamefont {P\'erez-Fern\'andez}\ \emph
  {et~al.}(2011)\citenamefont {P\'erez-Fern\'andez}, \citenamefont {Cejnar},
  \citenamefont {Arias}, \citenamefont {Dukelsky}, \citenamefont
  {Garc\'{i}a-Ramos},\ and\ \citenamefont {Rela\~no}}]{Fernandez2011}%
  \BibitemOpen
  \bibfield  {author} {\bibinfo {author} {\bibfnamefont {P.}~\bibnamefont
  {P\'erez-Fern\'andez}}, \bibinfo {author} {\bibfnamefont {P.}~\bibnamefont
  {Cejnar}}, \bibinfo {author} {\bibfnamefont {J.~M.}\ \bibnamefont {Arias}},
  \bibinfo {author} {\bibfnamefont {J.}~\bibnamefont {Dukelsky}}, \bibinfo
  {author} {\bibfnamefont {J.~E.}\ \bibnamefont {Garc\'{i}a-Ramos}}, \ and\
  \bibinfo {author} {\bibfnamefont {A.}~\bibnamefont {Rela\~no}},\ }\bibfield
  {title} {\enquote {\bibinfo {title} {Quantum quench influenced by an
  excited-state phase transition},}\ }\href {\doibase
  10.1103/PhysRevA.83.033802} {\bibfield  {journal} {\bibinfo  {journal} {Phys.
  Rev. A}\ }\textbf {\bibinfo {volume} {83}},\ \bibinfo {pages} {033802}
  (\bibinfo {year} {2011})}\BibitemShut {NoStop}%
\bibitem [{\citenamefont {Altland}\ and\ \citenamefont
  {Haake}(2012)}]{Altland2012PRL}%
  \BibitemOpen
  \bibfield  {author} {\bibinfo {author} {\bibfnamefont {Alexander}\
  \bibnamefont {Altland}}\ and\ \bibinfo {author} {\bibfnamefont {Fritz}\
  \bibnamefont {Haake}},\ }\bibfield  {title} {\enquote {\bibinfo {title}
  {Quantum chaos and effective thermalization},}\ }\href {\doibase
  10.1103/PhysRevLett.108.073601} {\bibfield  {journal} {\bibinfo  {journal}
  {Phys. Rev. Lett.}\ }\textbf {\bibinfo {volume} {108}},\ \bibinfo {pages}
  {073601} (\bibinfo {year} {2012})}\BibitemShut {NoStop}%
\bibitem [{\citenamefont {Shen}\ \emph {et~al.}(2017)\citenamefont {Shen},
  \citenamefont {Zhang}, \citenamefont {Fan},\ and\ \citenamefont
  {Zhai}}]{Shen2017}%
  \BibitemOpen
  \bibfield  {author} {\bibinfo {author} {\bibfnamefont {Huitao}\ \bibnamefont
  {Shen}}, \bibinfo {author} {\bibfnamefont {Pengfei}\ \bibnamefont {Zhang}},
  \bibinfo {author} {\bibfnamefont {Ruihua}\ \bibnamefont {Fan}}, \ and\
  \bibinfo {author} {\bibfnamefont {Hui}\ \bibnamefont {Zhai}},\ }\bibfield
  {title} {\enquote {\bibinfo {title} {Out-of-time-order correlation at a
  quantum phase transition},}\ }\href {\doibase 10.1103/PhysRevB.96.054503}
  {\bibfield  {journal} {\bibinfo  {journal} {Phys. Rev. B}\ }\textbf {\bibinfo
  {volume} {96}},\ \bibinfo {pages} {054503} (\bibinfo {year}
  {2017})}\BibitemShut {NoStop}%
\bibitem [{\citenamefont {Lerma-Hern\'andez}\ \emph {et~al.}(2018)\citenamefont
  {Lerma-Hern\'andez}, \citenamefont {Ch\'avez-Carlos}, \citenamefont
  {Bastarrachea-Magnani}, \citenamefont {Santos},\ and\ \citenamefont
  {Hirsch}}]{Lerma2018}%
  \BibitemOpen
  \bibfield  {author} {\bibinfo {author} {\bibfnamefont {Sergio}\ \bibnamefont
  {Lerma-Hern\'andez}}, \bibinfo {author} {\bibfnamefont {Jorge}\ \bibnamefont
  {Ch\'avez-Carlos}}, \bibinfo {author} {\bibfnamefont {Miguel~A.}\
  \bibnamefont {Bastarrachea-Magnani}}, \bibinfo {author} {\bibfnamefont
  {Lea~F.}\ \bibnamefont {Santos}}, \ and\ \bibinfo {author} {\bibfnamefont
  {Jorge~G.}\ \bibnamefont {Hirsch}},\ }\bibfield  {title} {\enquote {\bibinfo
  {title} {Analytical description of the survival probability of coherent
  states in regular regimes},}\ }\href
  {http://stacks.iop.org/1751-8121/51/i=47/a=475302} {\bibfield  {journal}
  {\bibinfo  {journal} {J. Phys. A}\ }\textbf {\bibinfo {volume} {51}},\
  \bibinfo {pages} {475302} (\bibinfo {year} {2018})}\BibitemShut {NoStop}%
\bibitem [{\citenamefont {Lerma-Hern\'andez}\ \emph {et~al.}(2019)\citenamefont
  {Lerma-Hern\'andez}, \citenamefont {Villase\~nor}, \citenamefont
  {Bastarrachea-Magnani}, \citenamefont {Torres-Herrera}, \citenamefont
  {Santos},\ and\ \citenamefont {Hirsch}}]{Lerma2019}%
  \BibitemOpen
  \bibfield  {author} {\bibinfo {author} {\bibfnamefont {S.}~\bibnamefont
  {Lerma-Hern\'andez}}, \bibinfo {author} {\bibfnamefont {D.}~\bibnamefont
  {Villase\~nor}}, \bibinfo {author} {\bibfnamefont {M.~A.}\ \bibnamefont
  {Bastarrachea-Magnani}}, \bibinfo {author} {\bibfnamefont {E.~J.}\
  \bibnamefont {Torres-Herrera}}, \bibinfo {author} {\bibfnamefont {L.~F.}\
  \bibnamefont {Santos}}, \ and\ \bibinfo {author} {\bibfnamefont {J.~G.}\
  \bibnamefont {Hirsch}},\ }\bibfield  {title} {\enquote {\bibinfo {title}
  {Dynamical signatures of quantum chaos and relaxation time scales in a
  spin-boson system},}\ }\href {\doibase 10.1103/PhysRevE.100.012218}
  {\bibfield  {journal} {\bibinfo  {journal} {Phys. Rev. E}\ }\textbf {\bibinfo
  {volume} {100}},\ \bibinfo {pages} {012218} (\bibinfo {year}
  {2019})}\BibitemShut {NoStop}%
\bibitem [{\citenamefont {Kloc}\ \emph {et~al.}(2018)\citenamefont {Kloc},
  \citenamefont {Str\'ansk\'y},\ and\ \citenamefont {Cejnar}}]{Kloc2018}%
  \BibitemOpen
  \bibfield  {author} {\bibinfo {author} {\bibfnamefont {Michal}\ \bibnamefont
  {Kloc}}, \bibinfo {author} {\bibfnamefont {Pavel}\ \bibnamefont
  {Str\'ansk\'y}}, \ and\ \bibinfo {author} {\bibfnamefont {Pavel}\
  \bibnamefont {Cejnar}},\ }\bibfield  {title} {\enquote {\bibinfo {title}
  {Quantum quench dynamics in {D}icke superradiance models},}\ }\href {\doibase
  10.1103/PhysRevA.98.013836} {\bibfield  {journal} {\bibinfo  {journal} {Phys.
  Rev. A}\ }\textbf {\bibinfo {volume} {98}},\ \bibinfo {pages} {013836}
  (\bibinfo {year} {2018})}\BibitemShut {NoStop}%
\bibitem [{\citenamefont {Kirton}\ \emph {et~al.}(2019)\citenamefont {Kirton},
  \citenamefont {Roses}, \citenamefont {Keeling},\ and\ \citenamefont
  {Dalla~Torre}}]{Kirton2019}%
  \BibitemOpen
  \bibfield  {author} {\bibinfo {author} {\bibfnamefont {Peter}\ \bibnamefont
  {Kirton}}, \bibinfo {author} {\bibfnamefont {Mor~M.}\ \bibnamefont {Roses}},
  \bibinfo {author} {\bibfnamefont {Jonathan}\ \bibnamefont {Keeling}}, \ and\
  \bibinfo {author} {\bibfnamefont {Emanuele~G.}\ \bibnamefont {Dalla~Torre}},\
  }\bibfield  {title} {\enquote {\bibinfo {title} {Introduction to the {D}icke
  model: From equilibrium to nonequilibrium, and vice versa},}\ }\href
  {\doibase 10.1002/qute.201800043} {\bibfield  {journal} {\bibinfo  {journal}
  {Adv. Quant. Tech.}\ }\textbf {\bibinfo {volume} {2}},\ \bibinfo {pages}
  {1800043} (\bibinfo {year} {2019})}\BibitemShut {NoStop}%
\bibitem [{\citenamefont {Villase{\~{n}}or}\ \emph {et~al.}(2020)\citenamefont
  {Villase{\~{n}}or}, \citenamefont {Pilatowsky-Cameo}, \citenamefont
  {Bastarrachea-Magnani}, \citenamefont {Lerma-Hern{\'{a}}ndez}, \citenamefont
  {Santos},\ and\ \citenamefont {Hirsch}}]{Villasenor2020}%
  \BibitemOpen
  \bibfield  {author} {\bibinfo {author} {\bibfnamefont {D}~\bibnamefont
  {Villase{\~{n}}or}}, \bibinfo {author} {\bibfnamefont {S}~\bibnamefont
  {Pilatowsky-Cameo}}, \bibinfo {author} {\bibfnamefont {M~A}\ \bibnamefont
  {Bastarrachea-Magnani}}, \bibinfo {author} {\bibfnamefont {S}~\bibnamefont
  {Lerma-Hern{\'{a}}ndez}}, \bibinfo {author} {\bibfnamefont {L~F}\
  \bibnamefont {Santos}}, \ and\ \bibinfo {author} {\bibfnamefont {J~G}\
  \bibnamefont {Hirsch}},\ }\bibfield  {title} {\enquote {\bibinfo {title}
  {Quantum vs classical dynamics in a spin-boson system: manifestations of
  spectral correlations and scarring},}\ }\href {\doibase
  10.1088/1367-2630/ab8ef8} {\bibfield  {journal} {\bibinfo  {journal} {New J.
  Phys.}\ }\textbf {\bibinfo {volume} {22}},\ \bibinfo {pages} {063036}
  (\bibinfo {year} {2020})}\BibitemShut {NoStop}%
\bibitem [{\citenamefont {Ch\'avez-Carlos}\ \emph {et~al.}(2019)\citenamefont
  {Ch\'avez-Carlos}, \citenamefont {L\'opez-del Carpio}, \citenamefont
  {Bastarrachea-Magnani}, \citenamefont {Str\'ansk\'y}, \citenamefont
  {Lerma-Hern\'andez}, \citenamefont {Santos},\ and\ \citenamefont
  {Hirsch}}]{Chavez2019}%
  \BibitemOpen
  \bibfield  {author} {\bibinfo {author} {\bibfnamefont {Jorge}\ \bibnamefont
  {Ch\'avez-Carlos}}, \bibinfo {author} {\bibfnamefont {B.}~\bibnamefont
  {L\'opez-del Carpio}}, \bibinfo {author} {\bibfnamefont {Miguel~A.}\
  \bibnamefont {Bastarrachea-Magnani}}, \bibinfo {author} {\bibfnamefont
  {Pavel}\ \bibnamefont {Str\'ansk\'y}}, \bibinfo {author} {\bibfnamefont
  {Sergio}\ \bibnamefont {Lerma-Hern\'andez}}, \bibinfo {author} {\bibfnamefont
  {Lea~F.}\ \bibnamefont {Santos}}, \ and\ \bibinfo {author} {\bibfnamefont
  {Jorge~G.}\ \bibnamefont {Hirsch}},\ }\bibfield  {title} {\enquote {\bibinfo
  {title} {Quantum and classical {L}yapunov exponents in atom-field interaction
  systems},}\ }\href {\doibase 10.1103/PhysRevLett.122.024101} {\bibfield
  {journal} {\bibinfo  {journal} {Phys. Rev. Lett.}\ }\textbf {\bibinfo
  {volume} {122}},\ \bibinfo {pages} {024101} (\bibinfo {year}
  {2019})}\BibitemShut {NoStop}%
\bibitem [{\citenamefont {Lewis-Swan}\ \emph {et~al.}(2019)\citenamefont
  {Lewis-Swan}, \citenamefont {Safavi-Naini}, \citenamefont {Bollinger},\ and\
  \citenamefont {Rey}}]{Lewis-Swan2019}%
  \BibitemOpen
  \bibfield  {author} {\bibinfo {author} {\bibfnamefont {R.~J.}\ \bibnamefont
  {Lewis-Swan}}, \bibinfo {author} {\bibfnamefont {A.}~\bibnamefont
  {Safavi-Naini}}, \bibinfo {author} {\bibfnamefont {J.~J.}\ \bibnamefont
  {Bollinger}}, \ and\ \bibinfo {author} {\bibfnamefont {A.~M.}\ \bibnamefont
  {Rey}},\ }\bibfield  {title} {\enquote {\bibinfo {title} {Unifying ,
  thermalization and entanglement through measurement of fidelity
  out-of-time-order correlators in the {D}icke model},}\ }\href {\doibase
  10.1038/s41467-019-09436-y} {\bibfield  {journal} {\bibinfo  {journal} {Nat.
  Comm.}\ }\textbf {\bibinfo {volume} {10}},\ \bibinfo {pages} {1581} (\bibinfo
  {year} {2019})}\BibitemShut {NoStop}%
\bibitem [{\citenamefont {Pilatowsky-Cameo}\ \emph
  {et~al.}(2020{\natexlab{a}})\citenamefont {Pilatowsky-Cameo}, \citenamefont
  {Ch\'avez-Carlos}, \citenamefont {Bastarrachea-Magnani}, \citenamefont
  {Str\'ansk\'y}, \citenamefont {Lerma-Hern\'andez}, \citenamefont {Santos},\
  and\ \citenamefont {Hirsch}}]{Pilatowsky2020}%
  \BibitemOpen
  \bibfield  {author} {\bibinfo {author} {\bibfnamefont {Sa\'ul}\ \bibnamefont
  {Pilatowsky-Cameo}}, \bibinfo {author} {\bibfnamefont {Jorge}\ \bibnamefont
  {Ch\'avez-Carlos}}, \bibinfo {author} {\bibfnamefont {Miguel~A.}\
  \bibnamefont {Bastarrachea-Magnani}}, \bibinfo {author} {\bibfnamefont
  {Pavel}\ \bibnamefont {Str\'ansk\'y}}, \bibinfo {author} {\bibfnamefont
  {Sergio}\ \bibnamefont {Lerma-Hern\'andez}}, \bibinfo {author} {\bibfnamefont
  {Lea~F.}\ \bibnamefont {Santos}}, \ and\ \bibinfo {author} {\bibfnamefont
  {Jorge~G.}\ \bibnamefont {Hirsch}},\ }\bibfield  {title} {\enquote {\bibinfo
  {title} {Positive quantum {L}yapunov exponents in experimental systems with a
  regular classical limit},}\ }\href {\doibase 10.1103/PhysRevE.101.010202}
  {\bibfield  {journal} {\bibinfo  {journal} {Phys. Rev. E}\ }\textbf {\bibinfo
  {volume} {101}},\ \bibinfo {pages} {010202(R)} (\bibinfo {year}
  {2020}{\natexlab{a}})}\BibitemShut {NoStop}%
\bibitem [{\citenamefont {De~Bernardis}\ \emph {et~al.}(2018)\citenamefont
  {De~Bernardis}, \citenamefont {Jaako},\ and\ \citenamefont
  {Rabl}}]{DeBernardis2018}%
  \BibitemOpen
  \bibfield  {author} {\bibinfo {author} {\bibfnamefont {Daniele}\ \bibnamefont
  {De~Bernardis}}, \bibinfo {author} {\bibfnamefont {Tuomas}\ \bibnamefont
  {Jaako}}, \ and\ \bibinfo {author} {\bibfnamefont {Peter}\ \bibnamefont
  {Rabl}},\ }\bibfield  {title} {\enquote {\bibinfo {title} {Cavity quantum
  electrodynamics in the nonperturbative regime},}\ }\href {\doibase
  10.1103/PhysRevA.97.043820} {\bibfield  {journal} {\bibinfo  {journal} {Phys.
  Rev. A}\ }\textbf {\bibinfo {volume} {97}},\ \bibinfo {pages} {043820}
  (\bibinfo {year} {2018})}\BibitemShut {NoStop}%
\bibitem [{\citenamefont {Frisk~Kockum}\ \emph {et~al.}(2019)\citenamefont
  {Frisk~Kockum}, \citenamefont {Miranowicz}, \citenamefont {De~Liberato},
  \citenamefont {Savasta},\ and\ \citenamefont {Nori}}]{Kockum2019}%
  \BibitemOpen
  \bibfield  {author} {\bibinfo {author} {\bibfnamefont {Anton}\ \bibnamefont
  {Frisk~Kockum}}, \bibinfo {author} {\bibfnamefont {Adam}\ \bibnamefont
  {Miranowicz}}, \bibinfo {author} {\bibfnamefont {Simone}\ \bibnamefont
  {De~Liberato}}, \bibinfo {author} {\bibfnamefont {Salvatore}\ \bibnamefont
  {Savasta}}, \ and\ \bibinfo {author} {\bibfnamefont {Franco}\ \bibnamefont
  {Nori}},\ }\bibfield  {title} {\enquote {\bibinfo {title} {Ultrastrong
  coupling between light and matter},}\ }\href@noop {} {\bibfield  {journal}
  {\bibinfo  {journal} {Nat. Rev. Phys.}\ }\textbf {\bibinfo {volume} {1}},\
  \bibinfo {pages} {19--40} (\bibinfo {year} {2019})}\BibitemShut {NoStop}%
\bibitem [{\citenamefont {Forn-D\'{\i}az}\ \emph {et~al.}(2019)\citenamefont
  {Forn-D\'{\i}az}, \citenamefont {Lamata}, \citenamefont {Rico}, \citenamefont
  {Kono},\ and\ \citenamefont {Solano}}]{FornDiaz2019}%
  \BibitemOpen
  \bibfield  {author} {\bibinfo {author} {\bibfnamefont {P.}~\bibnamefont
  {Forn-D\'{\i}az}}, \bibinfo {author} {\bibfnamefont {L.}~\bibnamefont
  {Lamata}}, \bibinfo {author} {\bibfnamefont {E.}~\bibnamefont {Rico}},
  \bibinfo {author} {\bibfnamefont {J.}~\bibnamefont {Kono}}, \ and\ \bibinfo
  {author} {\bibfnamefont {E.}~\bibnamefont {Solano}},\ }\bibfield  {title}
  {\enquote {\bibinfo {title} {Ultrastrong coupling regimes of light-matter
  interaction},}\ }\href {\doibase 10.1103/RevModPhys.91.025005} {\bibfield
  {journal} {\bibinfo  {journal} {Rev. Mod. Phys.}\ }\textbf {\bibinfo {volume}
  {91}},\ \bibinfo {pages} {025005} (\bibinfo {year} {2019})}\BibitemShut
  {NoStop}%
\bibitem [{\citenamefont {Le~Boit\'e}()}]{LeBoite2020}%
  \BibitemOpen
  \bibfield  {author} {\bibinfo {author} {\bibfnamefont {Alexandre}\
  \bibnamefont {Le~Boit\'e}},\ }\bibfield  {title} {\enquote {\bibinfo {title}
  {Theoretical methods for ultrastrong light–matter interactions},}\ }\href
  {\doibase 10.1002/qute.201900140} {\bibfield  {journal} {\bibinfo  {journal}
  {Adv. Quant. Tech.}\ }\textbf {\bibinfo {volume} {n/a}},\ \bibinfo {pages}
  {1900140}},\ \Eprint
  {http://arxiv.org/abs/https://onlinelibrary.wiley.com/doi/pdf/10.1002/qute.201900140}
  {https://onlinelibrary.wiley.com/doi/pdf/10.1002/qute.201900140} \BibitemShut
  {NoStop}%
\bibitem [{\citenamefont {Baden}\ \emph {et~al.}(2014)\citenamefont {Baden},
  \citenamefont {Arnold}, \citenamefont {Grimsmo}, \citenamefont {Parkins},\
  and\ \citenamefont {Barrett}}]{Baden2014}%
  \BibitemOpen
  \bibfield  {author} {\bibinfo {author} {\bibfnamefont {Markus~P.}\
  \bibnamefont {Baden}}, \bibinfo {author} {\bibfnamefont {Kyle~J.}\
  \bibnamefont {Arnold}}, \bibinfo {author} {\bibfnamefont {Arne~L.}\
  \bibnamefont {Grimsmo}}, \bibinfo {author} {\bibfnamefont {Scott}\
  \bibnamefont {Parkins}}, \ and\ \bibinfo {author} {\bibfnamefont {Murray~D.}\
  \bibnamefont {Barrett}},\ }\bibfield  {title} {\enquote {\bibinfo {title}
  {Realization of the {D}icke model using cavity-assisted {R}aman
  transitions},}\ }\href {\doibase 10.1103/PhysRevLett.113.020408} {\bibfield
  {journal} {\bibinfo  {journal} {Phys. Rev. Lett.}\ }\textbf {\bibinfo
  {volume} {113}},\ \bibinfo {pages} {020408} (\bibinfo {year}
  {2014})}\BibitemShut {NoStop}%
\bibitem [{\citenamefont {Zhang}\ \emph {et~al.}(2018)\citenamefont {Zhang},
  \citenamefont {Lee}, \citenamefont {Kumar}, \citenamefont {Arnold},
  \citenamefont {Masson}, \citenamefont {Grimsmo}, \citenamefont {Parkins},\
  and\ \citenamefont {Barrett}}]{Zhang2018}%
  \BibitemOpen
  \bibfield  {author} {\bibinfo {author} {\bibfnamefont {Zhiqiang}\
  \bibnamefont {Zhang}}, \bibinfo {author} {\bibfnamefont {Chern~Hui}\
  \bibnamefont {Lee}}, \bibinfo {author} {\bibfnamefont {Ravi}\ \bibnamefont
  {Kumar}}, \bibinfo {author} {\bibfnamefont {K.~J.}\ \bibnamefont {Arnold}},
  \bibinfo {author} {\bibfnamefont {Stuart~J.}\ \bibnamefont {Masson}},
  \bibinfo {author} {\bibfnamefont {A.~L.}\ \bibnamefont {Grimsmo}}, \bibinfo
  {author} {\bibfnamefont {A.~S.}\ \bibnamefont {Parkins}}, \ and\ \bibinfo
  {author} {\bibfnamefont {M.~D.}\ \bibnamefont {Barrett}},\ }\bibfield
  {title} {\enquote {\bibinfo {title} {Dicke-model simulation via
  cavity-assisted {R}aman transitions},}\ }\href {\doibase
  10.1103/PhysRevA.97.043858} {\bibfield  {journal} {\bibinfo  {journal} {Phys.
  Rev. A}\ }\textbf {\bibinfo {volume} {97}},\ \bibinfo {pages} {043858}
  (\bibinfo {year} {2018})}\BibitemShut {NoStop}%
\bibitem [{\citenamefont {Cohn}\ \emph {et~al.}(2018)\citenamefont {Cohn},
  \citenamefont {Safavi-Naini}, \citenamefont {Lewis-Swan}, \citenamefont
  {Bohnet}, \citenamefont {G\"arttner}, \citenamefont {Gilmore}, \citenamefont
  {Jordan}, \citenamefont {Rey}, \citenamefont {Bollinger},\ and\ \citenamefont
  {Freericks}}]{Cohn2018}%
  \BibitemOpen
  \bibfield  {author} {\bibinfo {author} {\bibfnamefont {J}~\bibnamefont
  {Cohn}}, \bibinfo {author} {\bibfnamefont {A}~\bibnamefont {Safavi-Naini}},
  \bibinfo {author} {\bibfnamefont {R~J}\ \bibnamefont {Lewis-Swan}}, \bibinfo
  {author} {\bibfnamefont {J~G}\ \bibnamefont {Bohnet}}, \bibinfo {author}
  {\bibfnamefont {M}~\bibnamefont {G\"arttner}}, \bibinfo {author}
  {\bibfnamefont {K~A}\ \bibnamefont {Gilmore}}, \bibinfo {author}
  {\bibfnamefont {J~E}\ \bibnamefont {Jordan}}, \bibinfo {author}
  {\bibfnamefont {A~M}\ \bibnamefont {Rey}}, \bibinfo {author} {\bibfnamefont
  {J~J}\ \bibnamefont {Bollinger}}, \ and\ \bibinfo {author} {\bibfnamefont
  {J~K}\ \bibnamefont {Freericks}},\ }\bibfield  {title} {\enquote {\bibinfo
  {title} {Bang-bang shortcut to adiabaticity in the {D}icke model as realized
  in a penning trap experiment},}\ }\href
  {http://stacks.iop.org/1367-2630/20/i=5/a=055013} {\bibfield  {journal}
  {\bibinfo  {journal} {New J. Phys.}\ }\textbf {\bibinfo {volume} {20}},\
  \bibinfo {pages} {055013} (\bibinfo {year} {2018})}\BibitemShut {NoStop}%
\bibitem [{\citenamefont {Safavi-Naini}\ \emph {et~al.}(2018)\citenamefont
  {Safavi-Naini}, \citenamefont {Lewis-Swan}, \citenamefont {Bohnet},
  \citenamefont {G\"arttner}, \citenamefont {Gilmore}, \citenamefont {Jordan},
  \citenamefont {Cohn}, \citenamefont {Freericks}, \citenamefont {Rey},\ and\
  \citenamefont {Bollinger}}]{Safavi2018}%
  \BibitemOpen
  \bibfield  {author} {\bibinfo {author} {\bibfnamefont {A.}~\bibnamefont
  {Safavi-Naini}}, \bibinfo {author} {\bibfnamefont {R.~J.}\ \bibnamefont
  {Lewis-Swan}}, \bibinfo {author} {\bibfnamefont {J.~G.}\ \bibnamefont
  {Bohnet}}, \bibinfo {author} {\bibfnamefont {M.}~\bibnamefont {G\"arttner}},
  \bibinfo {author} {\bibfnamefont {K.~A.}\ \bibnamefont {Gilmore}}, \bibinfo
  {author} {\bibfnamefont {J.~E.}\ \bibnamefont {Jordan}}, \bibinfo {author}
  {\bibfnamefont {J.}~\bibnamefont {Cohn}}, \bibinfo {author} {\bibfnamefont
  {J.~K.}\ \bibnamefont {Freericks}}, \bibinfo {author} {\bibfnamefont {A.~M.}\
  \bibnamefont {Rey}}, \ and\ \bibinfo {author} {\bibfnamefont {J.~J.}\
  \bibnamefont {Bollinger}},\ }\bibfield  {title} {\enquote {\bibinfo {title}
  {Verification of a many-ion simulator of the {D}icke model through slow
  quenches across a phase transition},}\ }\href {\doibase
  10.1103/PhysRevLett.121.040503} {\bibfield  {journal} {\bibinfo  {journal}
  {Phys. Rev. Lett.}\ }\textbf {\bibinfo {volume} {121}},\ \bibinfo {pages}
  {040503} (\bibinfo {year} {2018})}\BibitemShut {NoStop}%
\bibitem [{\citenamefont {Jaako}\ \emph {et~al.}(2016)\citenamefont {Jaako},
  \citenamefont {Xiang}, \citenamefont {Garcia-Ripoll},\ and\ \citenamefont
  {Rabl}}]{Jaako2016}%
  \BibitemOpen
  \bibfield  {author} {\bibinfo {author} {\bibfnamefont {Tuomas}\ \bibnamefont
  {Jaako}}, \bibinfo {author} {\bibfnamefont {Ze-Liang}\ \bibnamefont {Xiang}},
  \bibinfo {author} {\bibfnamefont {Juan~Jos\'e}\ \bibnamefont
  {Garcia-Ripoll}}, \ and\ \bibinfo {author} {\bibfnamefont {Peter}\
  \bibnamefont {Rabl}},\ }\bibfield  {title} {\enquote {\bibinfo {title}
  {Ultrastrong-coupling phenomena beyond the dicke model},}\ }\href {\doibase
  10.1103/PhysRevA.94.033850} {\bibfield  {journal} {\bibinfo  {journal} {Phys.
  Rev. A}\ }\textbf {\bibinfo {volume} {94}},\ \bibinfo {pages} {033850}
  (\bibinfo {year} {2016})}\BibitemShut {NoStop}%
\bibitem [{\citenamefont {Lewenkopf}\ \emph {et~al.}(1991)\citenamefont
  {Lewenkopf}, \citenamefont {Nemes}, \citenamefont {Marvulle}, \citenamefont
  {Pato},\ and\ \citenamefont {Wreszinski}}]{Lewenkopf1991}%
  \BibitemOpen
  \bibfield  {author} {\bibinfo {author} {\bibfnamefont {C.H}\ \bibnamefont
  {Lewenkopf}}, \bibinfo {author} {\bibfnamefont {M.C}\ \bibnamefont {Nemes}},
  \bibinfo {author} {\bibfnamefont {V}~\bibnamefont {Marvulle}}, \bibinfo
  {author} {\bibfnamefont {M.P}\ \bibnamefont {Pato}}, \ and\ \bibinfo {author}
  {\bibfnamefont {W.F}\ \bibnamefont {Wreszinski}},\ }\bibfield  {title}
  {\enquote {\bibinfo {title} {Level statistics transitions in the spin-boson
  model},}\ }\href {\doibase https://doi.org/10.1016/0375-9601(91)90575-S}
  {\bibfield  {journal} {\bibinfo  {journal} {Phys. Lett. A}\ }\textbf
  {\bibinfo {volume} {155}},\ \bibinfo {pages} {113 -- 116} (\bibinfo {year}
  {1991})}\BibitemShut {NoStop}%
\bibitem [{\citenamefont {Emary}\ and\ \citenamefont
  {Brandes}(2003{\natexlab{b}})}]{Emary2003PRL}%
  \BibitemOpen
  \bibfield  {author} {\bibinfo {author} {\bibfnamefont {Clive}\ \bibnamefont
  {Emary}}\ and\ \bibinfo {author} {\bibfnamefont {Tobias}\ \bibnamefont
  {Brandes}},\ }\bibfield  {title} {\enquote {\bibinfo {title} {Quantum chaos
  triggered by precursors of a quantum phase transition: The {D}icke model},}\
  }\href {\doibase 10.1103/PhysRevLett.90.044101} {\bibfield  {journal}
  {\bibinfo  {journal} {Phys. Rev. Lett.}\ }\textbf {\bibinfo {volume} {90}},\
  \bibinfo {pages} {044101} (\bibinfo {year} {2003}{\natexlab{b}})}\BibitemShut
  {NoStop}%
\bibitem [{\citenamefont {Bastarrachea-Magnani}\ \emph
  {et~al.}(2014{\natexlab{a}})\citenamefont {Bastarrachea-Magnani},
  \citenamefont {Lerma-Hern\'andez},\ and\ \citenamefont
  {Hirsch}}]{Bastarrachea2014b}%
  \BibitemOpen
  \bibfield  {author} {\bibinfo {author} {\bibfnamefont {M.~A.}\ \bibnamefont
  {Bastarrachea-Magnani}}, \bibinfo {author} {\bibfnamefont {S.}~\bibnamefont
  {Lerma-Hern\'andez}}, \ and\ \bibinfo {author} {\bibfnamefont {J.~G.}\
  \bibnamefont {Hirsch}},\ }\bibfield  {title} {\enquote {\bibinfo {title}
  {Comparative quantum and semiclassical analysis of atom-field systems. ii.
  {C}haos and regularity},}\ }\href {\doibase 10.1103/PhysRevA.89.032102}
  {\bibfield  {journal} {\bibinfo  {journal} {Phys. Rev. A}\ }\textbf {\bibinfo
  {volume} {89}},\ \bibinfo {pages} {032102} (\bibinfo {year}
  {2014}{\natexlab{a}})}\BibitemShut {NoStop}%
\bibitem [{\citenamefont {Bastarrachea-Magnani}\ \emph
  {et~al.}(2015)\citenamefont {Bastarrachea-Magnani}, \citenamefont {del
  Carpio}, \citenamefont {Lerma-Hern\'andez},\ and\ \citenamefont
  {Hirsch}}]{Bastarrachea2015}%
  \BibitemOpen
  \bibfield  {author} {\bibinfo {author} {\bibfnamefont {Miguel~Angel}\
  \bibnamefont {Bastarrachea-Magnani}}, \bibinfo {author} {\bibfnamefont
  {Baldemar~L\'opez}\ \bibnamefont {del Carpio}}, \bibinfo {author}
  {\bibfnamefont {Sergio}\ \bibnamefont {Lerma-Hern\'andez}}, \ and\ \bibinfo
  {author} {\bibfnamefont {Jorge~G}\ \bibnamefont {Hirsch}},\ }\bibfield
  {title} {\enquote {\bibinfo {title} {Chaos in the {D}icke model: quantum and
  semiclassical analysis},}\ }\href
  {http://stacks.iop.org/1402-4896/90/i=6/a=068015} {\bibfield  {journal}
  {\bibinfo  {journal} {Phys. Scripta}\ }\textbf {\bibinfo {volume} {90}},\
  \bibinfo {pages} {068015} (\bibinfo {year} {2015})}\BibitemShut {NoStop}%
\bibitem [{\citenamefont {Bastarrachea-Magnani}\ \emph
  {et~al.}(2016)\citenamefont {Bastarrachea-Magnani}, \citenamefont
  {L\'opez-del{-}Carpio}, \citenamefont {Ch\'avez-Carlos}, \citenamefont
  {Lerma-Hern\'andez},\ and\ \citenamefont {Hirsch}}]{Bastarrachea2016PRE}%
  \BibitemOpen
  \bibfield  {author} {\bibinfo {author} {\bibfnamefont {M.~A.}\ \bibnamefont
  {Bastarrachea-Magnani}}, \bibinfo {author} {\bibfnamefont {B.}~\bibnamefont
  {L\'opez-del{-}Carpio}}, \bibinfo {author} {\bibfnamefont {J.}~\bibnamefont
  {Ch\'avez-Carlos}}, \bibinfo {author} {\bibfnamefont {S.}~\bibnamefont
  {Lerma-Hern\'andez}}, \ and\ \bibinfo {author} {\bibfnamefont {J.~G.}\
  \bibnamefont {Hirsch}},\ }\bibfield  {title} {\enquote {\bibinfo {title}
  {Delocalization and quantum chaos in atom-field systems},}\ }\href {\doibase
  10.1103/PhysRevE.93.022215} {\bibfield  {journal} {\bibinfo  {journal} {Phys.
  Rev. E}\ }\textbf {\bibinfo {volume} {93}},\ \bibinfo {pages} {022215}
  (\bibinfo {year} {2016})}\BibitemShut {NoStop}%
\bibitem [{\citenamefont {Ch\'avez-Carlos}\ \emph {et~al.}(2016)\citenamefont
  {Ch\'avez-Carlos}, \citenamefont {Bastarrachea-Magnani}, \citenamefont
  {Lerma-Hern\'andez},\ and\ \citenamefont {Hirsch}}]{Chavez2016}%
  \BibitemOpen
  \bibfield  {author} {\bibinfo {author} {\bibfnamefont {J.}~\bibnamefont
  {Ch\'avez-Carlos}}, \bibinfo {author} {\bibfnamefont {M.~A.}\ \bibnamefont
  {Bastarrachea-Magnani}}, \bibinfo {author} {\bibfnamefont {S.}~\bibnamefont
  {Lerma-Hern\'andez}}, \ and\ \bibinfo {author} {\bibfnamefont {J.~G.}\
  \bibnamefont {Hirsch}},\ }\bibfield  {title} {\enquote {\bibinfo {title}
  {Classical chaos in atom-field systems},}\ }\href {\doibase
  10.1103/PhysRevE.94.022209} {\bibfield  {journal} {\bibinfo  {journal} {Phys.
  Rev. E}\ }\textbf {\bibinfo {volume} {94}},\ \bibinfo {pages} {022209}
  (\bibinfo {year} {2016})}\BibitemShut {NoStop}%
\bibitem [{\citenamefont {de~Aguiar}\ \emph {et~al.}(1991)\citenamefont
  {de~Aguiar}, \citenamefont {Furuya}, \citenamefont {Lewenkopf},\ and\
  \citenamefont {Nemes}}]{Deaguiar1991}%
  \BibitemOpen
  \bibfield  {author} {\bibinfo {author} {\bibfnamefont {M.~A.~M.}\
  \bibnamefont {de~Aguiar}}, \bibinfo {author} {\bibfnamefont {K.}~\bibnamefont
  {Furuya}}, \bibinfo {author} {\bibfnamefont {C.~H.}\ \bibnamefont
  {Lewenkopf}}, \ and\ \bibinfo {author} {\bibfnamefont {M.~C.}\ \bibnamefont
  {Nemes}},\ }\bibfield  {title} {\enquote {\bibinfo {title} {Particle-spin
  coupling in a chaotic system: Localization-delocalization in the {H}usimi
  distributions},}\ }\href {http://stacks.iop.org/0295-5075/15/i=2/a=003}
  {\bibfield  {journal} {\bibinfo  {journal} {EPL (Europhys. Lett.)}\ }\textbf
  {\bibinfo {volume} {15}},\ \bibinfo {pages} {125} (\bibinfo {year}
  {1991})}\BibitemShut {NoStop}%
\bibitem [{\citenamefont {Pilatowsky-Cameo}\ \emph
  {et~al.}(2020{\natexlab{b}})\citenamefont {Pilatowsky-Cameo}, \citenamefont
  {Villase{\~{n}}or}, \citenamefont {Bastarrachea-Magnani}, \citenamefont
  {Lerma-Hern\'andez}, \citenamefont {Santos},\ and\ \citenamefont
  {Hirsch}}]{PilatowskyARXIV}%
  \BibitemOpen
  \bibfield  {author} {\bibinfo {author} {\bibfnamefont {Sa\'ul}\ \bibnamefont
  {Pilatowsky-Cameo}}, \bibinfo {author} {\bibfnamefont {David}\ \bibnamefont
  {Villase{\~{n}}or}}, \bibinfo {author} {\bibfnamefont {Miguel~A.}\
  \bibnamefont {Bastarrachea-Magnani}}, \bibinfo {author} {\bibfnamefont
  {Sergio}\ \bibnamefont {Lerma-Hern\'andez}}, \bibinfo {author} {\bibfnamefont
  {Lea~F.}\ \bibnamefont {Santos}}, \ and\ \bibinfo {author} {\bibfnamefont
  {Jorge~G.}\ \bibnamefont {Hirsch}},\ }\href@noop {} {\enquote {\bibinfo
  {title} {Does scarring prevent ergodicity?}}\ } (\bibinfo {year}
  {2020}{\natexlab{b}}),\ \Eprint {http://arxiv.org/abs/2009.00626}
  {arXiv:2009.00626 [cond-mat.stat-mech]} \BibitemShut {NoStop}%
\bibitem [{\citenamefont {Wang}\ and\ \citenamefont {Robnik}(2020)}]{Wang2020}%
  \BibitemOpen
  \bibfield  {author} {\bibinfo {author} {\bibfnamefont {Qian}\ \bibnamefont
  {Wang}}\ and\ \bibinfo {author} {\bibfnamefont {Marko}\ \bibnamefont
  {Robnik}},\ }\bibfield  {title} {\enquote {\bibinfo {title} {{Statistical
  properties of the localization measure of chaotic eigenstates in the Dicke
  model}},}\ }\href {\doibase 10.1103/PhysRevE.102.032212} {\bibfield
  {journal} {\bibinfo  {journal} {Phys. Rev. E}\ }\textbf {\bibinfo {volume}
  {102}},\ \bibinfo {pages} {032212} (\bibinfo {year} {2020})}\BibitemShut
  {NoStop}%
\bibitem [{foo({\natexlab{a}})}]{footExcept}%
  \BibitemOpen
  \href@noop {} {} \bibinfo {note} {{We note that chaotic
  systems may exhibit small stable islands, which are exceptions to this
  picture.}}\BibitemShut {Stop}%
\bibitem [{\citenamefont {Tomiya}\ \emph {et~al.}(2019)\citenamefont {Tomiya},
  \citenamefont {Sakamoto},\ and\ \citenamefont {Heller}}]{Tomiya2019}%
  \BibitemOpen
  \bibfield  {author} {\bibinfo {author} {\bibfnamefont {Mitsuyoshi}\
  \bibnamefont {Tomiya}}, \bibinfo {author} {\bibfnamefont {Shoichi}\
  \bibnamefont {Sakamoto}}, \ and\ \bibinfo {author} {\bibfnamefont {Eric~J.}\
  \bibnamefont {Heller}},\ }\bibfield  {title} {\enquote {\bibinfo {title}
  {Periodic orbit scar in wavepacket propagation},}\ }\href {\doibase
  10.1142/S0129183119500268} {\bibfield  {journal} {\bibinfo  {journal} {Int.
  J. Mod. Phys. C}\ }\textbf {\bibinfo {volume} {30}},\ \bibinfo {pages}
  {1950026} (\bibinfo {year} {2019})}\BibitemShut {NoStop}%
\bibitem [{\citenamefont {Bastarrachea-Magnani}\ \emph
  {et~al.}(2014{\natexlab{b}})\citenamefont {Bastarrachea-Magnani},
  \citenamefont {Lerma-Hern\'andez},\ and\ \citenamefont
  {Hirsch}}]{Bastarrachea2014a}%
  \BibitemOpen
  \bibfield  {author} {\bibinfo {author} {\bibfnamefont {M.~A.}\ \bibnamefont
  {Bastarrachea-Magnani}}, \bibinfo {author} {\bibfnamefont {S.}~\bibnamefont
  {Lerma-Hern\'andez}}, \ and\ \bibinfo {author} {\bibfnamefont {J.~G.}\
  \bibnamefont {Hirsch}},\ }\bibfield  {title} {\enquote {\bibinfo {title}
  {Comparative quantum and semiclassical analysis of atom-field systems. {I}.
  {D}ensity of states and excited-state quantum phase transitions},}\ }\href
  {\doibase 10.1103/PhysRevA.89.032101} {\bibfield  {journal} {\bibinfo
  {journal} {Phys. Rev. A}\ }\textbf {\bibinfo {volume} {89}},\ \bibinfo
  {pages} {032101} (\bibinfo {year} {2014}{\natexlab{b}})}\BibitemShut
  {NoStop}%
\bibitem [{\citenamefont {Ribeiro}\ \emph {et~al.}(2006)\citenamefont
  {Ribeiro}, \citenamefont {de~Aguiar},\ and\ \citenamefont
  {de~Toledo~Piza}}]{Ribeiro2006}%
  \BibitemOpen
  \bibfield  {author} {\bibinfo {author} {\bibfnamefont {A.~D.}\ \bibnamefont
  {Ribeiro}}, \bibinfo {author} {\bibfnamefont {M.~A.~M.}\ \bibnamefont
  {de~Aguiar}}, \ and\ \bibinfo {author} {\bibfnamefont {A.~F.~R.}\
  \bibnamefont {de~Toledo~Piza}},\ }\bibfield  {title} {\enquote {\bibinfo
  {title} {The semiclassical coherent state propagator for systems with
  spin},}\ }\href {http://stacks.iop.org/0305-4470/39/i=12/a=016} {\bibfield
  {journal} {\bibinfo  {journal} {J. Phys. A}\ }\textbf {\bibinfo {volume}
  {39}},\ \bibinfo {pages} {3085} (\bibinfo {year} {2006})}\BibitemShut
  {NoStop}%
\bibitem [{\citenamefont {de~Aguiar}\ and\ \citenamefont
  {Malta}(1988)}]{DeAguiar1988}%
  \BibitemOpen
  \bibfield  {author} {\bibinfo {author} {\bibfnamefont {M.A.M.}\ \bibnamefont
  {de~Aguiar}}\ and\ \bibinfo {author} {\bibfnamefont {C.P.}\ \bibnamefont
  {Malta}},\ }\bibfield  {title} {\enquote {\bibinfo {title} {Isochronous and
  period doubling bifurcations of periodic solutions of non-integrable
  {H}amiltonian systems with reflexion symmetries},}\ }\href {\doibase
  10.1016/0167-2789(88)90029-2} {\bibfield  {journal} {\bibinfo  {journal}
  {Phys. D}\ }\textbf {\bibinfo {volume} {30}},\ \bibinfo {pages} {413--424}
  (\bibinfo {year} {1988})}\BibitemShut {NoStop}%
\bibitem [{\citenamefont {Weinstein}(1973)}]{Weinstein1973}%
  \BibitemOpen
  \bibfield  {author} {\bibinfo {author} {\bibfnamefont {Alan}\ \bibnamefont
  {Weinstein}},\ }\bibfield  {title} {\enquote {\bibinfo {title} {Normal modes
  for nonlinear hamiltonian systems},}\ }\href {\doibase 10.1007/bf01405263}
  {\bibfield  {journal} {\bibinfo  {journal} {Inv. Math.}\ }\textbf {\bibinfo
  {volume} {20}},\ \bibinfo {pages} {47--57} (\bibinfo {year}
  {1973})}\BibitemShut {NoStop}%
\bibitem [{foo({\natexlab{b}})}]{footNote}%
  \BibitemOpen
  \href@noop {} {} \bibinfo {note} {{The parity symmetry of
  the model implies that
  $\mathcal{Q}_{k}(q,p;Q,P)$=$\mathcal{Q}_{k}(-q,p;-Q,P)$ for all eigenstates
  $\ket{E_k}$. Thus, $\mathcal{P}^{A}_{k} =
  2\mathcal{P}(\mathcal{O}_{\epsilon_k}^A,\hat{\rho}_k)
  =2\mathcal{P}(\widetilde{\mathcal{O}}_{\epsilon_k}^A,\hat{\rho}_k)$ and
  $\mathcal{P}^{B}_k=2\mathcal{P}(\mathcal{O}_{\epsilon_k}^B,\hat{\rho}_k)
  =2\mathcal{P}(\widetilde{\mathcal{O}}_{\epsilon_k}^B,\hat{\rho}_k)$.}}\BibitemShut
  {Stop}%
\bibitem [{\citenamefont {Peres}(1984)}]{Peres1984PRL}%
  \BibitemOpen
  \bibfield  {author} {\bibinfo {author} {\bibfnamefont {Asher}\ \bibnamefont
  {Peres}},\ }\bibfield  {title} {\enquote {\bibinfo {title} {New conserved
  quantities and test for regular spectra},}\ }\href {\doibase
  10.1103/PhysRevLett.53.1711} {\bibfield  {journal} {\bibinfo  {journal}
  {Phys. Rev. Lett.}\ }\textbf {\bibinfo {volume} {53}},\ \bibinfo {pages}
  {1711--1713} (\bibinfo {year} {1984})}\BibitemShut {NoStop}%
\bibitem [{\citenamefont {Gutzwiller}(1971)}]{Gutzwiller1971}%
  \BibitemOpen
  \bibfield  {author} {\bibinfo {author} {\bibfnamefont {Martin~C.}\
  \bibnamefont {Gutzwiller}},\ }\bibfield  {title} {\enquote {\bibinfo {title}
  {Periodic orbits and classical quantization conditions},}\ }\href {\doibase
  10.1063/1.1665596} {\bibfield  {journal} {\bibinfo  {journal} {J. of Math.
  Phys.}\ }\textbf {\bibinfo {volume} {12}},\ \bibinfo {pages} {343--358}
  (\bibinfo {year} {1971})}\BibitemShut {NoStop}%
\bibitem [{\citenamefont {Schliemann}(2015)}]{Schliemann2015}%
  \BibitemOpen
  \bibfield  {author} {\bibinfo {author} {\bibfnamefont {John}\ \bibnamefont
  {Schliemann}},\ }\bibfield  {title} {\enquote {\bibinfo {title} {Coherent
  quantum dynamics: What fluctuations can tell},}\ }\href {\doibase
  10.1103/PhysRevA.92.022108} {\bibfield  {journal} {\bibinfo  {journal} {Phys.
  Rev. A}\ }\textbf {\bibinfo {volume} {92}},\ \bibinfo {pages} {022108}
  (\bibinfo {year} {2015})}\BibitemShut {NoStop}%
\bibitem [{\citenamefont {Furuya}\ \emph {et~al.}(1992)\citenamefont {Furuya},
  \citenamefont {de~Aguiar}, \citenamefont {Lewenkopf},\ and\ \citenamefont
  {Nemes}}]{Furuya1992}%
  \BibitemOpen
  \bibfield  {author} {\bibinfo {author} {\bibfnamefont {K}~\bibnamefont
  {Furuya}}, \bibinfo {author} {\bibfnamefont {M.A.M}\ \bibnamefont
  {de~Aguiar}}, \bibinfo {author} {\bibfnamefont {C.H}\ \bibnamefont
  {Lewenkopf}}, \ and\ \bibinfo {author} {\bibfnamefont {M.C}\ \bibnamefont
  {Nemes}},\ }\bibfield  {title} {\enquote {\bibinfo {title} {{H}usimi
  distributions of a spin-boson system and the signatures of its classical
  dynamics},}\ }\href {\doibase 10.1016/0003-4916(92)90179-p} {\bibfield
  {journal} {\bibinfo  {journal} {Ann. of Phys.}\ }\textbf {\bibinfo {volume}
  {216}},\ \bibinfo {pages} {313--322} (\bibinfo {year} {1992})}\BibitemShut
  {NoStop}%
\bibitem [{\citenamefont {Baranger}\ \emph {et~al.}(1988)\citenamefont
  {Baranger}, \citenamefont {Davies},\ and\ \citenamefont
  {Mahoney}}]{Baranger1988}%
  \BibitemOpen
  \bibfield  {author} {\bibinfo {author} {\bibfnamefont {M.}~\bibnamefont
  {Baranger}}, \bibinfo {author} {\bibfnamefont {K.T.R.}\ \bibnamefont
  {Davies}}, \ and\ \bibinfo {author} {\bibfnamefont {J.H.}\ \bibnamefont
  {Mahoney}},\ }\bibfield  {title} {\enquote {\bibinfo {title} {The calculation
  of periodic trajectories},}\ }\href {\doibase 10.1016/s0003-4916(88)80018-6}
  {\bibfield  {journal} {\bibinfo  {journal} {Ann. of Phys.}\ }\textbf
  {\bibinfo {volume} {186}},\ \bibinfo {pages} {95--110} (\bibinfo {year}
  {1988})}\BibitemShut {NoStop}%
\bibitem [{\citenamefont {Simonovi{\'{c}}}(1999)}]{Simonovi1999}%
  \BibitemOpen
  \bibfield  {author} {\bibinfo {author} {\bibfnamefont {N.~S.}\ \bibnamefont
  {Simonovi{\'{c}}}},\ }\bibfield  {title} {\enquote {\bibinfo {title}
  {Calculations of periodic orbits: The monodromy method and application to
  regularized systems},}\ }\href {\doibase 10.1063/1.166457} {\bibfield
  {journal} {\bibinfo  {journal} {Chaos}\ }\textbf {\bibinfo {volume} {9}},\
  \bibinfo {pages} {854--864} (\bibinfo {year} {1999})}\BibitemShut {NoStop}%
\bibitem [{\citenamefont {Gaspard}(1998)}]{Gaspard1998}%
  \BibitemOpen
  \bibfield  {author} {\bibinfo {author} {\bibfnamefont {Pierre}\ \bibnamefont
  {Gaspard}},\ }\href@noop {} {\emph {\bibinfo {title} {Chaos, scattering and
  statistical mechanics}}}\ (\bibinfo  {publisher} {Cambridge Univ. Press},\
  \bibinfo {year} {1998})\BibitemShut {NoStop}%
\end{thebibliography}
%


\end{document}